\documentclass[12pt,web,draftcls,onecolumn]{ieeecolor}
\usepackage{generic}

\usepackage{amsmath,amssymb,amsfonts,amsthm}
\usepackage{pmat}
\usepackage{array}
\usepackage{mathrsfs}
\usepackage{setspace} % change the size of table 
\usepackage{arydshln}
\usepackage[framemethod=tikz]{mdframed}
\usepackage{multirow}
\usepackage{multicol}
\usepackage{diagbox}
\usepackage{hyperref}

\usepackage{subcaption}
\usepackage{graphicx}                        %图形宏包
\usepackage{epsfig}                          %使用EPS图片
\usepackage{epstopdf}                        %EPS图片转PDF格式
\usepackage{amssymb}                         %支持EPS图片
\usepackage{color}                           %插入颜色

\usepackage[justification=centering]{caption} %图片居中

\theoremstyle{plain}\newtheorem{theorem}{Theorem}
\theoremstyle{plain}\newtheorem{lemma}{Lemma}
\theoremstyle{plain}
\theoremstyle{plain}\newtheorem{proposition}{Proposition}
\theoremstyle{definition}
\theoremstyle{definition}\newtheorem{remark}{Remark}
\theoremstyle{definition}\newtheorem{definition}{Definition}
\theoremstyle{definition}
\theoremstyle{definition}

\def\BibTeX{{\rm B\kern-.05em{\sc i\kern-.025em b}\kern-.08em
    T\kern-.1667em\lower.7ex\hbox{E}\kern-.125emX}}
\begin{document}
\title{Frequency-dependent Switching Control for Disturbance Attenuation of Linear Systems}
\author{Jingjing Zhang, Jan Heiland, Peter Benner, Xin Du 
\thanks{This work was supported in part by the National Natural Science Foundation of China under Grant (No. 61873336) and International Corporation Project of Shanghai Science and Technology Commission under Grant 21190780300, in part by The National Key Research and Development Program (No. 2020YFB1708200) , in part by 111 Project (No. D18003) granted by the State Administration of Foreign Experts Affairs (SAFEA), in part by "International Cooperative Training Project for Innovative Talents (autonomous intelligent system)" granted by China Scholarship Council and undertaken by Shanghai University}
\thanks{Jingjing Zhang is with the School of Mechatronic Engineering and Automation, Shanghai University, Shanghai, 200444, China,  and also with the Max Planck Institute for Dynamics of Complex Technical Systems, 39106 Magdeburg, Germany (e-mail: zhang@mpi-magdeburg.mpg.de) }
\thanks{Jan Heiland is with the Max Planck Institute for Dynamics of Complex Technical Systems, 39106 Magdeburg, Germany, and also with the Faculty of Mathematics, Otto von Guericke University Magdeburg, 39106 Magdeburg, Germany (e-mail: heiland@mpi-magdeburg.mpg.de)}
\thanks{Peter Benner is with the Max Planck Institute for Dynamics of Complex Technical Systems, 39106 Magdeburg, Germany, and also with the Faculty of Mathematics, Otto von Guericke University Magdeburg, 39106 Magdeburg, Germany (e-mail: benner@mpi-magdeburg.mpg.de).}
\thanks{Xin Du is with the  School of Mechatronic Engineering and Automation, Shanghai University, Shanghai, 200444, China, and also with the Shanghai Key Laboratory of Power Station Automation Technology, Shanghai University, Shanghai, 200444, China. Corresponding author. (e-mail: duxin@shu.edu.cn)}}

\maketitle

\begin{abstract}
 The generalized Kalman–Yakubovich–Popov lemma as established by Iwasaki and Hara in 2005 marks a milestone in the analysis and synthesis of linear systems from a  finite-frequency perspective. Given a pre-specified frequency band, it allows us to produce passive controllers with excellent in-band disturbance attenuation performance at the expense of some of the out-of-band performance. This paper focuses on control design of linear systems in the presence of disturbances with non-strictly or non-stationary limited frequency spectrum.  We first propose a class of frequency-dependent excited energy functions (FD-EEF) as well as frequency-dependent excited power functions (FD-EPF), which possess a desirable frequency-selectiveness property with regard to the in-band and out-of-band excited energy as well as excited power of the system.  Based upon a group of frequency-selective passive controllers, we then develop a frequency-dependent switching control (FDSC) scheme that
 selects the most appropriate controller at runtime. We show that our FDSC scheme is capable to approximate the solid in-band performance while maintaining acceptable out-of-band performance with regard to global time horizons as well as localized time horizons. The method is illustrated by a
 commonly used benchmark model.
\end{abstract}


\section{Introduction}
\label{sec:introduction}
\IEEEPARstart {D}{isturbance} attenuation is one of the most important problems in control theory and engineering practice, which can be traced back to 1940s and has attracted increasing attention since the 1980s. A controller with fixed gain based on $H_\infty$ norm probably was firstly designed by George Zames \cite{fdsczames1981feedback}. Thanks to the outstanding contributions from J. William Helton, Doyle John and other pioneers \cite{francis1987linear}-\cite{Iwasaki1994Automatica},  $H_\infty$ control theory was established and recognized as a standard control design framework for disturbance attenuation. The most appealing feature of $H_\infty$ control is the {\it worst-case} optimality in presence of arbitrary $L_2$ bounded disturbances, which makes it become the most appropriate scheme if there does not exist {\it a priori} knowledge regarding the disturbance signal.

From a frequency-domain perspective, $H_\infty$ control intrinsically treats the component of disturbance at any frequency point equally. Due to this property, $H_\infty$ control can be viewed as an indispensable tool for dealing with disturbance of uniform or quasi-uniform spectrum over the full infinite frequency range, such as {\it white noise} or {\it pseudo white noise}. In practice, the  spectrum of disturbances may only span a limited frequency range, or the dominating components of the spectrum may focus on a limited frequency range. For example, seismic wave signals causing building vibration are focused on the frequency range from 0.3Hz to 8.8Hz \cite{fdscChen2010finite}, the critical source effecting the positioning accuracy of a hard disk drives servo system is the high-frequency narrow-band disturbance between 8kHz and 10kHz \cite{du2006generalized}. A fundamental tool for focusing the considerations on predefined frequency ranges is provided by the generalized Kalman-Yakubovich-Popov (gKYP) lemma as it has been developed from the standard KYP lemma by Iwasaki and Hara \cite{fdscIwasaki2005generalized, fdscIwasaki2005time} in 2005. With the aid of gKYP lemma, the framework of $H_\infty$ control theory was extended from the entire-frequency setting to the finite-frequency setting \cite{fdsciwasaki2007feedback, hara2006robust, iwasaki2004robust}. The gKYP lemma based control design method is capable to produce passive controllers that obviously improve the in-band disturbance attenuation performance, while sacrificing the out-of-band performance. Many successful applications of gKYP lemma based control design were reported in the last decade \cite{fdscSun2020advanced, zhang2021ride, zhu2015active, tan2013integrated, graham2007analysis, berrada2018sliding, wei2011sensor, xu2018reliable, liu2020innovative, preda2018robust, du2007generalized, ikegami2018numerical, yao2021research, zhou2013research}. 

The gKYP lemma based control design methods achieve in-band performance improvement by minimizing the in-band maximum singular value of the closed-loop transfer function. On the other side, the out-of-band maximum singular value of the closed-loop transfer function generally becomes very large at a certain frequencies (or frequency ranges).  In many applications, however, the assumption that the frequency spectrum of the disturbances are strictly limited is an oversimplification and non-negligible out-of-band  disturbances are very likely to occur \cite{wei2011sensor, xu2018reliable}. Besides, due to the change of system operational conditions, the dominating frequency range may occasionally switch from the supposed one to another. For example, the road roughness is the main source causing vehicle vibration \cite{zhang2021ride}. Normally a vehicle moves along flat roads, the roughness-induced disturbance varies slowly so that it exhibits low-frequency (LF) dominance. However, high-frequency (HF) components will be the dominating part when the vehicle passes through severely uneven road segments. In smart grid systems, the statistical data often show that the fluctuation of wind power, photovoltaic power as well as load may take a fast and slow separated feature over different time windows in hourly or daily scale \cite{ikegami2018numerical, yao2021research, zhou2013research}. Correspondingly, the dominating frequency components of the disturbance may be of significant difference with regard to moving time windows. In those cases, the overall disturbance  attenuation capability of the gKYP lemma designed passive controller may significantly deteriorate even in the presence of tiny out-of-band components. In particular, it may lead to disastrous and unacceptable system response over certain time slots due to the non-stationary of the dominating frequency band. One possible solution is to enlarge the frequency range or introduce multiple frequency ranges in the gKYP lemma based control design, however, the in-band performance decays quickly in this manner.  
 
In this paper, we develop appropriate improvements of the gKYP lemma based passive control design method. Inspired by the time-domain interpretation of the gKYP lemma \cite{fdscIwasaki2005time}, we first introduce a class of {\it frequency-dependent excited energy functions (FD-EEF)} and {\it frequency-dependent excited power functions (FD-EPF)} with respect to the system state and its derivative. We unveil the frequency-selectiveness property of  {\it FD-EEF} and {\it FD-EPF} on indicating the density of accumulated energy over moving time windows, as well as instantaneous power excited by in-band and out-of-band components of the disturbance. Based upon the frequency-selectiveness property, we propose a {\it frequency-dependent switching control (FDSC)} scheme orchestrating an in-band and a group of out-of-band oriented passive controllers, which are generated by the gKYP lemma based control design while setting different frequency ranges. Moreover, a performance analysis of the closed-loop system implemented with FDSC is carried out through an extensive re-exploitation of the gKYP lemma.  It is shown that the proposed FDSC scheme provides an appealing balance between the in-band performance and out-of-band performance. To a great extent, the solid in-band performance of the in-band oriented passive controller can be inherited, while the effects caused by out-of-band components are well counteracted by the out-of-band oriented passive controllers. In particular, the proposed FDSC scheme preserves good asymptotic performance while avoiding intolerable transient system responses during the out-of-band components dominated short-term time phases.   

The rest of this paper is organized as follows. Section II starts by revisiting the gKYP lemma based control design.  Section III introduces the concept of FD-EEF and FD-EPF and investigates its frequency-selectiveness property. The mechanism of the FDSC scheme as well as the switching logic design are presented in Section IV. Simulation results based on a generic and a common benchmark model are illustrated in Section VI to verify the effectiveness and advantages of the proposed FDSC control scheme. Section VII concludes this paper with some remarks.

Notations: For a matrix $M$, $M^{T}$, $M^{*}$ and $M^{\bot}$ means its transpose, complex conjugate transpose and orthogonal complement, respectively. $M<(>)0$ and $M\leq(\geq)0$ mean the matrix $M$ is negative (positive) and semi-negative (semi-positive) definite, respectively. For matrices $\Phi$ and $P$, $\Phi\otimes P$ represents their Kronecker product.  $\Omega_e:=(-\infty,\infty)$ defines the entire frequency range. Given a finite frequency range $\Omega$, $\bar{\Omega}:=\Omega_e\setminus \Omega$ represents its complement.

\section{Preliminaries and Problem Statements}
\subsection{Frequency-limited signals and frequency-dominated signals}
We briefly recall the connection between a signal $d \colon [0,\infty) \to \mathbb R$ on the time-domain and its representation 
 \[\mathcal D (\omega) =\int_{0} ^{\infty} d(t) e^{-\jmath\omega t},\]
 in the frequency-domain as it is obtained by the Fourier Transform (FT) under the tacit assumption that $d(t)$ is square integrable. The inverse FT recovers $d(t)$ from its frequency representation $\mathcal D (\omega)$ via
\begin{equation}
		d(t)=\frac{1}{2\pi}\int_{-\infty}^{+\infty} {\mathcal D(\omega)e^{\jmath\omega t}d\omega}.
\end{equation}
Furthermore, we note that if $\mathcal D (\omega)=0$ for $\omega\notin \Omega_i$, then 
\begin{equation}
		d(t)=\frac{1}{2\pi}\sum\limits_{i=1}^I \int_{\Omega_{i}} {\mathcal D(\omega)e^{\jmath\omega t}d\omega},   
\end{equation}
where $\Omega_i,i=1,2,\ldots,I$ are finite frequency ranges.

\begin{mdframed}[hidealllines=true,backgroundcolor=red!4,innerleftmargin=4pt,innerrightmargin=4pt,leftmargin=1pt,rightmargin=1pt]
\begin{definition}(Frequency-Limited Signals)\label{definition1}
\end{definition} 
\end{mdframed}
\vspace{-6mm}
\begin{mdframed}[hidealllines=true,backgroundcolor=red!1,innerleftmargin=4pt,innerrightmargin=4pt,leftmargin=1pt,rightmargin=1pt]
A signal $d(t)$ will be referred as \emph{frequency-limited with respect to a pre-specified finite frequency range $\Omega_i$}, if it spectrum is zero beyond $\Omega_i$, i.e.,
\begin{equation}\label{fre-lim-signal}
\mathcal D (\omega) =0 , ~\forall \omega \notin \Omega_i.
 \end{equation}
 \end{mdframed}
\noindent Generally, a finite-frequency range $\Omega_i$ is defined within the category of low-frequency (LF), or middle-frequency (MF), or high-frequency (HF). Specifically, LF, MF, HF are finite frequency ranges which can be explicitly described as 
\begin{equation*}
	\begin{split}
	&\Omega_i:= \Omega_l= [-\varpi_l, +\varpi_l],  {\kern 92pt}{\rm  LF~ case},  \\
	&\Omega_i:=\Omega_m =[-\varpi_2, -\varpi_1] \cup [+\varpi_1, +\varpi_2],  {\kern 9pt}{\rm MF~ case},\\
	&\Omega_i:= \Omega_h=(-\infty, -\varpi_h] \cup [\varpi_h, +\infty) ,   {\kern 24pt} {\rm HF~ case}.
	\end{split}
\end{equation*}

\begin{mdframed}[hidealllines=true,backgroundcolor=red!4,innerleftmargin=4pt,innerrightmargin=4pt,leftmargin=1pt,rightmargin=1pt]
\begin{definition}(Frequency-Dominated Signals)\label{definition2}
\end{definition} 
\end{mdframed}
\vspace{-6mm}
\begin{mdframed}[hidealllines=true,backgroundcolor=red!1,innerleftmargin=4pt,innerrightmargin=4pt,leftmargin=1pt,rightmargin=1pt]
A signal $d(t)$ will be referred as \emph{frequency-dominated with respect to a pre-specified finite frequency range $\Omega_i$}, if the energy of in-band and out-of-band components satisfies
\begin{equation}\label{eqn_FD-EEF1}
\int_{{\Omega _{i}}} {\mathcal D^*(\jmath\omega) \mathcal D(\jmath\omega)d\omega} > \int_{{\bar \Omega _{i}}} {\mathcal D^*(\jmath\omega)\mathcal D(\jmath\omega)d\omega}.
\end{equation}
Furthermore,  the ratio between the energy of in-band components and the total energy
\[
\alpha(\Omega_i)  = \frac{\int_{{\Omega _{i}}} {\mathcal D^*(\jmath\omega) \mathcal D (\jmath\omega)d\omega} }{\int_{{\Omega_e} } {\mathcal D^*(\jmath\omega) \mathcal D(\jmath\omega)d\omega} },
\]
will be referred as the \emph{dominance degree} of $\Omega_i$.
 \end{mdframed}

\noindent  Frequency-dominance provides a more realistic index to characterize the importance of a concerned frequency range. In fact, frequency-limited signals can be viewed as the extreme case of frequency-dominance with $\alpha(\Omega_i)=1$. 

Noting that the properties in Definition \ref{definition1} and \ref{definition2} are both presented from an infinite time horizon perspective. If we turns to a localized time horizon viewpoint, a frequency-dominated signal may exhibit more complicated characteristics. The dominance degree with respect to the given frequency range may vary slowly over a long time window or dramatically over a short time slot. One way to reveal the time-localized behavior of a given non-stationary signal is to use the well-known Short Time Fourier Transform (STFT). The essential idea of STFT is to perform the FT on each segmented narrow time interval of the total time series to find out instantaneous frequency spectra. In this paper, to present the non-stationary frequency dominance of a given signal, we resort to the notion of time-localized {\it pseudo} frequency spectrum.  

\begin{mdframed}[hidealllines=true,backgroundcolor=red!4,innerleftmargin=4pt,innerrightmargin=4pt,leftmargin=1pt,rightmargin=1pt]
\begin{definition}(Time-Localized {\it Pseudo} Frequency Spectrum)\label{definition3}
\end{definition} 
\end{mdframed}
\vspace{-6mm}
\begin{mdframed}[hidealllines=true,backgroundcolor=red!1,innerleftmargin=4pt,innerrightmargin=4pt,leftmargin=1pt,rightmargin=1pt]
A function $\mathcal D_{T_l}(\omega)$ will be referred as \emph{time-localized pseudo frequency spectrum} of $d(t)$ over time interval $T_l:[t_l,t_{l+1})(l=1,2,\ldots)$, if $d(t)$ over $T_l$ can be recovered from $\mathcal D_{T_l}(\omega)$ by the following inverse FT
\begin{equation}
\begin{array}{l}
d(t)=\frac{1}{2\pi} \int_{0}^{+\infty} {\mathcal D_{T_l}(\omega)e^{\jmath\omega t}d\omega}, {\kern 5pt}t \in T_l.
\end{array}
 \end{equation}  
 \end{mdframed}

\noindent With the time-localized {\it pseudo} frequency spectrum in mind, a signal $d(t)$ can be expressed in a joint time-frequency manner as follows:
\[d(t)=\left\{ \begin{array}{l}
\sum\limits_{i=i}^I \int_{\Omega_{i}} {\mathcal D_{T_1}(\jmath\omega)e^{\jmath\omega t}d\omega} , \forall t \in T_1:[0,t_1),\\ 
\sum\limits_{i=1}^I \int_{\Omega_{i}} {\mathcal D_{T_2}(\jmath\omega)e^{\jmath\omega t}d\omega} , \forall t \in T_2:[t_1,t_2),\\ 
\ldots\\
\sum\limits_{i=1}^I \int_{\Omega_{i}} {\mathcal D_{T_l}(\jmath\omega)e^{\jmath\omega t}d\omega} , \forall t \in T_l:[t_{l-1},t_l),\\
 \ldots,\\
\end{array} \right.\]
where $\{T_l, T_2,\ldots,T_l,\ldots \}$ are successive segmented time slots, $\{\Omega_1, \Omega_2,\ldots,\Omega_I\}$ are non-overlapping frequency ranges. Correspondingly, a group of indices $\alpha(\Omega_i, T_l) $ can be generated as:
\[
\alpha(\Omega_i, T_l)  = \frac{\int_{{\Omega _{i}}} {\mathcal D_{T_l}^*(\jmath\omega)\mathcal D_{T_l}(\jmath\omega)d\omega} }{\int_{{\Omega_e} } {\mathcal D_{T_l}^*(\jmath\omega)\mathcal D_{T_l}(\jmath\omega)d\omega} },\\ 
\]
and used to describe the time-localized frequency dominance level of the in-band frequency components. Furthermore, in case that the time-localized frequency dominance level is varying over time, i.e., $\alpha(\Omega_i, T_l) \neq \alpha(\Omega_i, T_r), l\neq r$, then $d(t)$ will be referred as non-stationary frequency-dominated signal with respect to a pre-specified finite frequency range $\Omega_{i}$. In practice, the frequency components of a signal may mainly fall into a frequency range $\Omega_j$ over a short-time window, while itself is a globally frequency-dominated signal with respect to $\Omega_i$, where $i\neq j$ and $\Omega_i \cap \Omega_j =\emptyset$. 

\subsection {Revisit of gKYP lemma based passive controller design}
Consider the following linear time-invariant system
\begin{equation}\label{eqn_plant}
\begin{split}
\dot{x}(t) &= Ax(t)+Bd(t),\\
z(t) &= C x(t) + Dd(t),
\end{split}
\end{equation}
where $x(t) \in {\mathbb{R}^n}$ is the state vector, $d(t) \in {\mathbb{R}^{n_d}}$ the disturbance vector and $z(t) \in {\mathbb{R}^{n_z}}$ the regulated output. Let $A$, $B$, $C$, and $D$ be real constant matrices with appropriate dimensions. To make the paper self-contained, we briefly include the $L_2$-gain analysis oriented form of generalized Kalman-Yakubovich-Popov (gKYP) lemma as it has been developed by Iwasaki and Hara to deal with finite-frequency analysis in controller synthesis:

\begin{mdframed}[hidealllines=true,backgroundcolor=red!4,innerleftmargin=4pt,innerrightmargin=4pt,leftmargin=1pt,rightmargin=1pt]
\begin{lemma}Generalized Kalman-Yakubovich-Popov (gKYP) Lemma 
\end{lemma} 
\end{mdframed}
\vspace{-6mm}
\begin{mdframed}[hidealllines=true,backgroundcolor=red!1,innerleftmargin=4pt,innerrightmargin=4pt,leftmargin=1pt,rightmargin=1pt]
Given a linear system \eqref{eqn_plant}, the following two statements are equivalent.\\
(1) For a finite frequency range $\Omega_i$, there exist matrices $P=P^*, Q=Q^*>0$ such that the following linear matrix inequality holds
\begin{equation}
\label{gKYP_LMI}
\begin{array}{l}
 \pmatset{1}{0.1pt}
  \pmatset{0}{0.1pt}
  \pmatset{2}{4pt}
  \pmatset{3}{4pt}
  \pmatset{4}{4pt}
  \pmatset{5}{4pt}
  \pmatset{6}{1pt}
  {\kern 10pt}
\begin{pmat}[{.}]
    A   & {\kern 4pt} B \cr
   I  &{\kern 4pt} 0 \cr
  \end{pmat}^*
(\Phi  \otimes P +\Psi \otimes Q) 
 \begin{pmat}[{.}]
    A   & {\kern 4pt}  B \cr
   I  &{\kern 4pt} 0 \cr
  \end{pmat} + \begin{pmat}[{.}]
   C   & {\kern 4pt}  D  \cr
    0   & {\kern 4pt} I \cr
  \end{pmat}^*
  \begin{pmat}[{.}]
   I   & 0  \cr
   0   & -{\gamma_{f}}^2 \cr
  \end{pmat} \begin{pmat}[{.}]
     C    & {\kern 4pt} D  \cr
    0   & {\kern 4pt}I \cr
  \end{pmat}
 \end{array}<0,
\end{equation}
where $\Psi$ are frequency variables, and 
\[\begin{array}{l}
 \Psi=  \pmatset{1}{0.1pt}
  \pmatset{0}{0.1pt}
  \pmatset{2}{4pt}
  \pmatset{3}{4pt}
  \pmatset{4}{4pt}
  \pmatset{5}{4pt}
  \pmatset{6}{1pt}
\begin{pmat}[{.}]
    -1   &  0   \cr
 0   &   \varpi^2_l \cr
  \end{pmat},     {\kern 63pt} {\rm if}~ \Omega_i=\Omega_l  \\
 \Psi= 
 \pmatset{1}{0.1pt}
  \pmatset{0}{0.1pt}
  \pmatset{2}{4pt}
  \pmatset{3}{4pt}
  \pmatset{4}{4pt}
  \pmatset{5}{4pt}
  \pmatset{6}{1pt}
\begin{pmat}[{.}]
    -1   &   \jmath \varpi_c   \cr
 -\jmath \varpi_c  &    -\varpi_1\varpi_2 \cr
  \end{pmat} ,    {\kern 28pt}{\rm if} ~\Omega_i=\Omega_m \\
  \Psi=  \pmatset{1}{0.1pt}
  \pmatset{0}{0.1pt}
  \pmatset{2}{4pt}
  \pmatset{3}{4pt}
  \pmatset{4}{4pt}
  \pmatset{5}{4pt}
  \pmatset{6}{1pt}
\begin{pmat}[{.}]
  1   &  0   \cr
 0   &  -\varpi^2_h  \cr
  \end{pmat},     {\kern 63pt}  {\rm if}~ \Omega_i=\Omega_h.
 \end{array}
\]
(2)  If the input signal $d(t)$ is frequency-limited with respect to $\Omega_i$, the  input-output gain in the $L_2$-norm sense satisfies 
       \begin{equation}\label{eq_gain_inf2}
        \frac{{{ \int_{0}^{+\infty} z^*(t)z(t)dt }}}{{\int_{0}^{+\infty}d^*(t)d(t)dt}}<\gamma_{i},
       \end{equation}
where $\gamma_i$ is a positive scalar. 
\end{mdframed}

\noindent Based on the gKYP lemma, routines for seeking sub-optimal gains of a passive controller which attenuates a frequency-limited disturbance were developed \cite{fdsciwasaki2007feedback, hara2006robust, iwasaki2004robust}. Given a finite frequency range $\Omega_{i}$, these methods define a sub-optimal passive state feedback law 
\begin{equation}
	u(t)=K_ix(t),
\end{equation}
so that for the linear system
\begin{equation}\label{con_plant}
	\begin{split}
		\dot{x}(t) &= Ax(t)+B_1d(t)+B_2u(t),\\
		z(t) &= C x(t) + D_1 d(t)+D_2u(t),
	\end{split}
\end{equation}
with a control input $u$ and the regulated output $z$ (and $A$, $B_1$, $B_2$, $C$, $D_{1}$ and $D_{2}$ are real constant matrices with appropriate dimensions), the transfer function $\mathbf G_i \colon d\to z$ of the corresponding closed-loop system satisfies
 \begin{equation}
	 \mathop{\sigma_{max}}(\mathbf  G_i (\jmath\omega)) < \gamma_i(\Omega_{i},K_i),  {\kern 5pt}\forall \omega\in \Omega_i
\end{equation}
with $\gamma_i(\Omega_{i},K_i)$ specified by the gKYP lemma.

\subsection{Problem statement}
Consider the linear system (\ref{con_plant}) in the presence of frequency-dominated disturbance $d(t)$. From an infinite time horizon perspective, we assume the dominating frequency range $\Omega^{in}$ is known {\it a priori} while its dominance degree $\alpha(\Omega^{in})$ is sufficiently large but unknown, i.e., $\alpha(\Omega^{in})\to 1$. From the viewpoint of localized time horizon, we assume the dominance degree $\alpha(\Omega^{in}, T_{l})$ could be stationary or non-stationary. In particular, we assume that an arbitrary small time-localized dominance degree over a time interval or intervals (i.e., $\alpha(\Omega^{in}, T_{x}) \to 0$) is an admissible scenarios in our problem setting. 

Based upon the gKYP lemma, an in-band passive state feedback controller 
      \begin{equation} 
      \label{in-band controller}
       u(t)=K^{in} x(t)
       \end{equation}
can be generated and assumed to be available in our design. To combat with the out-of-band components, a heuristic way is to assign a group of concerned frequency ranges $\{\Omega^{out}_i|\Omega^{out}_i\cap \Omega^{in}=\emptyset,  i=1,2,\ldots,I\}$ and produce a group of  state feedback passive controllers
       \begin{equation} 
          \label{out-band controller}
          u(t)=K^{out}_i x(t),~i=1,2,\ldots,I 
       \end{equation}
also by applying the gKYP lemma based control design method in the first stage, and then facilitate those candidates in a proper way. 

Based on the in-band as well as the out-of-band passive controllers, the control objective of this work is to develop an adaptive mechanism which can utilize the passive controllers such that: 
\begin{mdframed}[hidealllines=true,backgroundcolor=red!1,innerleftmargin=14pt,
innerrightmargin=14pt,leftmargin=12pt,rightmargin=12pt]
\begin{itemize}
\item[(P.1)] On the global disturbance attenuation performance: \\
          The adaptive mechanism inherits the solid global disturbance attenuation performance over infinite time horizon of the in-band controller (\ref{in-band controller}) as close as possible
       \begin{equation}
       \frac{{{ \int_{0}^{\infty}z^*(t)z(t)dt }}}{{\int_{0}^{\infty}d^*(t)d(t)dt}}\rightarrow\gamma(\Omega^{in}, K^{in}).
       \end{equation}
\end{itemize}
\begin{itemize}
\item[(P.2)] On the localized disturbance attenuation performance: \\
                      The adaptive mechanism ensures an acceptable disturbance attenuation performance over an arbitrarily selected time window $T_l$
       \begin{equation}
       \frac{{{ \int_{T_l} z^*(t)z(t)dt }}}{{\int_{T_l}d^*(t)d(t)dt}}< \gamma_{tol},
       \end{equation}
   where $\gamma_{tol}$ generally is set up as a smaller value of  $\gamma(\Omega_i^{out}, K^{in})$.
\end{itemize}
\end{mdframed}

\section{Frequency-dependent switching control}
\subsection{Frequency-selective functions} 
Beside the input-output analysis, sometimes it is of great importance to observe the response of system states under the excitation of the input signal $d(t)$. From the state space equation of linear system \eqref{eqn_plant}, we have 
\begin{equation}
x(t):=e^{At}x(0)+e^{At} \int_{0}^{t}  e^{-A \tau} B d(\tau)d\tau, ~~t \in[0, +\infty),
\end{equation}
where $x(0)$ is the initial condition. Generally, one could use the following {\it excited energy function} to measure the excitation degree of the system caused by input signal $d(t)$ over an infinite time horizon, i.e.,
\begin{equation}\label{eef}
 \mathbb S(Q)= \int_{0}^{+\infty}  x^*(t) Q  x(t)dt,
\end{equation}
where $Q=Q^*>0$ is a user-defined matrix. The {\it excited energy function} reduces to the well-known {\it controllability} or {\it observability} Gramian by setting $Q=I$. Furthermore, to measure the excitation degree by input signal $d(t)$ over a given finite time horizon $T_l: [t_l, t_{l+1})$, we define the {\it excited energy function} as 
\begin{equation}
 \mathbb S(Q, T_l)= \int_{t_l}^{t_{l+1}}  x^*(t) Q  x(t)dt,
\end{equation}
in which the lower and upper bounds of the integral is replaced by the starting and ending time instants of the given time window, respectively.  

\begin{mdframed}[hidealllines=true,backgroundcolor=red!4,innerleftmargin=4pt,innerrightmargin=4pt,leftmargin=1pt,rightmargin=1pt]
\begin{definition}[ Frequency-Dependent Excited Energy Function (FD-EEF)]
\end{definition}
\end{mdframed}
\vspace{-6mm}
\begin{mdframed}[hidealllines=true,backgroundcolor=red!1,innerleftmargin=4pt,innerrightmargin=4pt,leftmargin=1pt,rightmargin=1pt]
\begin{itemize} 
\item[$\rhd$] \underline{Infinite time horizon case}. Given a linear system \eqref{eqn_plant} with input signal $d(t)$, a finite-frequency range $\Omega_{i}$ and a positive definite matrix  $Q_i$,  the function 
\begin{equation}\label{eqn_FD-EEF1}
 \mathbb S(\Omega_i, Q_i)= \int_{0}^{+\infty}[\dot x^*(t) \kern 4pt  x^*(t)] (\Psi_i \otimes Q_i) [ \dot x^*(t) \kern 4pt x^*(t)]^*dt
\end{equation}
will be referred to as {\it frequency-dependent excited energy function (FD-EEF)} of the linear system \eqref{eqn_plant}  with respect to $(\Omega_i, Q_i)$.
\end{itemize}

\begin{itemize}
\item[$\rhd$] \underline{localized-time horizon case}. Given a linear system \eqref{eqn_plant} with input signal $d(t)$, a finite-frequency range $\Omega_i$  and a positive definite matrix  $Q_i$,  the function 
\begin{equation}\label{eqn_FD-EEF1}
 \mathbb S(\Omega_i, Q_i, T_l)= \int_{t_l}^{t_{l+1}}[\dot x^*(t) \kern 4pt  x^*(t)] (\Psi_i \otimes Q_i) [ \dot x^*(t) \kern 4pt x^*(t)]^*dt
\end{equation}
will be referred to as {\it frequency-dependent excited energy function (FD-EEF)} of the linear system \eqref{eqn_plant} with respect to $(\Omega_i, Q_i)$ over the time window $T_l$. 
\end{itemize}
\end{mdframed}

 \begin{mdframed}[hidealllines=true,backgroundcolor=red!4,innerleftmargin=4pt,
 innerrightmargin=4pt,leftmargin=1pt,rightmargin=1pt]
\begin{proposition} (Frequency-Selectiveness of FD-EEF)
\end{proposition}
\end{mdframed}
\vspace{-6mm}
 \begin{mdframed}[hidealllines=true,backgroundcolor=red!1,innerleftmargin=4pt,
 innerrightmargin=4pt,leftmargin=1pt,rightmargin=1pt]
FD-EEF exhibits a frequency-dependent selectiveness property with respect to the frequency spectrum of the system state $x(t)$, specifically,  
\begin{itemize}
\item[1)]  \underline{Infinite time horizon case}. Suppose that $x(t)$ has a frequency-limited  spectrum $\mathscr X(\jmath\omega)$ over the infinite time horizon with respect to $\Omega_i$. For given frequency range $\Omega^{in}$ satisfying $\Omega^{in}\supset \Omega_i$ and  frequency range $\Omega^{out}$ satisfying $\Omega^{out}\cap\Omega_i=\emptyset$,  we have \begin{subequations}
   \begin{align}
&  \mathbb S(\Omega^{in}, Q^{in})>0, \\
&  \mathbb S(\Omega^{out}, Q^{out})<0.
   \end{align}
\end{subequations}
where $Q^{in}$ and $Q^{out}$ are positive definite matrices. 
\end{itemize}
\begin{itemize}
\item[2)]\underline{Localized-time horizon case}. Suppose that $x(t)$ has a frequency-limited {\it pseudo} spectrum $\mathscr X_{T_l}(\jmath\omega)$ over a finite time horizon $T_l$ with respect to $\Omega_i$, For given frequency range $\Omega^{in}$ satisfying $\Omega^{in}\supset \Omega_i$ and frequency range $\Omega^{out}$ satisfying $\Omega^{out}\cap\Omega_i=\emptyset$,  we have  \begin{subequations}
   \begin{align}
& \mathbb S(\Omega^{in}, Q^{in},T_l)>0, \\
& \mathbb S(\Omega^{out}, Q^{out},T_l)<0.
   \end{align}
\end{subequations}
\end{itemize}
 \end{mdframed}

 \noindent {\textbf {Proof}}.  1)~\underline{Infinite time horizon case}.  According to Parseval's theorem and inverse FT, the equivalent formulation of FD-EEF in frequency-domain is

\begin{spacing}{1.0}
\begin{subequations}\label{KYPlurD1}
   \begin{align}
&   \mathscr F^{-1}( \mathbb S(\Omega^{in}, Q^{in}))=\frac{1}{{2\pi }}\int_{-\infty}^{+\infty}  [ (\jmath \omega)^* {\kern 2pt}  1] \Psi^{in} [(\jmath \omega)^*  {\kern 2pt} 1]^* X^*(\jmath\omega)Q^{in}X(\jmath\omega)d\omega,   \\
&  \mathscr F^{-1}( \mathbb S(\Omega^{out}, Q^{out}))=\frac{1}{{2\pi }}\int_{-\infty}^{+\infty}  [ (\jmath \omega)^* {\kern 2pt}  1] \Psi^{out} [(\jmath \omega)^*  {\kern 2pt} 1]^* X^*(\jmath\omega)Q^{out}X(\jmath\omega)d\omega.
   \end{align}
\end{subequations}
\end{spacing}
\noindent For a given matrix $\Psi_i$ characterizing a frequency range $\Omega_i$, we have  
\[  [ (\jmath \omega)^* {\kern 2pt}  1]\Psi_f [(\jmath \omega)^*  {\kern 2pt} 1]^* = \left\{ \begin{array}{l}
( \varpi_l- \jmath \omega)(\varpi_l+\jmath\omega) >0,  {\kern 10pt}{\rm if} ~ \Omega_{i}  \subset \Omega_l \\
( \omega-\varpi_1)(\varpi_2- \omega)>0, {\kern 15pt}{\rm if} ~\Omega_{i} \subset \Omega_m \\
(\jmath\omega- \varpi_h)(\jmath \omega+\varpi_h)>0, {\kern 5pt} {\rm if} ~\Omega_{i} \subset \Omega_h.
\end{array} \right.\]
Noting the relationship between $\Omega^{in}, \Omega^{out}$ and $\Omega_i$, we have
\begin{subequations}
   \begin{align}
& [  (\jmath \omega)^* {\kern 2pt}  1] \Psi^{in} [(\jmath \omega)^*  {\kern 2pt} 1]^*\ge 0, \forall \omega \in \Omega_i,\\
& [ (\jmath \omega)^* {\kern 2pt}  1] \Psi^{out} [(\jmath \omega)^*  {\kern 2pt} 1]^*\le 0, \forall \omega \notin \Omega_i.
   \end{align}
\end{subequations}
\noindent With the frequency-limited assumption imposed on $x(t)$, it is easy to derive that
\begin{spacing}{1.0}
\begin{subequations}\label{KYPlurD2}
   \begin{align}
&   \mathscr F^{-1}( \mathbb S(\Omega^{in}, Q^{in}))>0,   \\
&  \mathscr F^{-1}( \mathbb S(\Omega^{out}, Q^{out}))<0. 
   \end{align}
\end{subequations}
\end{spacing}
 2)~The counterpart to prove the \underline{localized-time horizon case} can be easily obtained by following the same routine.

\begin{remark}
{\it If $x(t)$ is not a strictly frequency-limited signal, then the positiveness or negativeness of FD-EEF can not be guaranteed. However, with the following in-band and out-of-band decomposition
\begin{equation}
 \mathscr F^{-1}( \mathbb S(\Omega_i, Q_i))= \mathscr F^{-1}( \mathbb S^{in}(\Omega_{i}, Q_i,)+ \mathscr F^{-1}( \mathbb S^{out}(\Omega_i, Q_i)),
\end{equation}
where
$$  \mathscr F^{-1}( \mathbb S^{in}(\Omega_i, Q_i))=\frac{1}{{2\pi }}\int_{\omega\in \Omega_i} [  (\jmath \omega)^* {\kern 2pt}  1] \Psi_i [(\jmath \omega)^*  {\kern 2pt} 1]^* X^*(\jmath\omega)Q_i X(\jmath\omega)d\omega >0,$$
$$ \mathscr F^{-1}( \mathbb S^{out}(\Omega_i, Q_i))=\frac{1}{{2\pi }}\int_{\omega \notin \Omega_i} [  (\jmath \omega)^* {\kern 2pt}  1] \Psi_i [(\jmath \omega)^*  {\kern 2pt} 1]^* X^*(\jmath\omega)Q_i X(\jmath\omega)d\omega <0,$$
the value of FD-EEF could reveal the dominance degree of a given frequency range $\Omega_i$. Specifically speaking, a positive value of FD-EEF means that the in-band excited energy is stronger than the out-of-band excited energy and vice versa. Furthermore, the larger value of FD-EEF we have, the larger dominance degree with respect to $\Omega_i$ can be concluded.  }
\end{remark}
 
\begin{remark} 
{\it The FD-EEF $\mathbb{S}(\Omega_i, Q_i)$ can be viewed as frequency-weighted extension of the standard excited energy function $\mathbb{S}(Q)$  \eqref{eef} , which can be equivalently interpreted in frequency-domain as follows:
\begin{equation}
 \mathscr F^{-1}( \mathbb S(Q))=\frac{1}{{2\pi }}\int_{-\infty}^{+\infty}  X^*(\jmath\omega)QX(\jmath\omega)d\omega,
\end{equation}
according to Parseval's theorem and inverse FT. While the standard excited energy function $\mathbb{S}(Q)$ measures the sum of energy excited by all the frequency components in a uniform way, the FD-EEF $\mathbb{S}(\Omega_i, Q_i)$ treats the in-band components with positive weighting factors and introduces negative weighting factors to the out-of-band components.  }
\end{remark}

In conjunction with the concept of energy, power is another crucial index to measure the strength of signals. Physically, power is defined as the amount of energy consumed per unit time. On the basis of  FD-EEF, we define a group of {\it frequency-dependent  excited power functions} as follows:
\begin{mdframed}[hidealllines=true,backgroundcolor=red!2,innerleftmargin=8pt,
innerrightmargin=8pt,leftmargin=1pt,rightmargin=1pt]
\begin{definition}(Frequency-Dependent Excited Power Function (FD-EPF))
\end{definition} 
\end{mdframed}
\vspace{-6mm}
\begin{mdframed}[hidealllines=true,backgroundcolor=red!1,innerleftmargin=8pt,
innerrightmargin=8pt,leftmargin=1pt,rightmargin=1pt]
\begin{itemize}
\item[1)] 
Given a linear system \eqref{eqn_plant},  the function 
\begin{equation}
 \mathbb P_a(\Omega_i, Q_i,T_l)= \mathbb S(\Omega_i, Q_i, T_l)/(t_{l+1}-t_l)
\end{equation}
 will be referred to as the {\it {\textbf {average}} finite frequency-dependent excited power function} with respect to the finite-frequency range $\Omega_i$ and the time window $T_l:[t_1, t_{l+1})$. 
\end{itemize}
\begin{itemize}
\item[2)] 
Given a linear system \eqref{eqn_plant},  the function 
\begin{equation}
 \mathbb P_t(\Omega_i, Q_i, t)=  [\dot x^*(t) \kern 4pt  x^*(t)] (\Psi_i \otimes Q_i) [ \dot x^*(t) \kern 4pt x^*(t)]^*
\end{equation}
as the {\it {\textbf {instantaneous}} finite frequency-dependent excited power function} of the linear system \eqref{eqn_plant}. 
\end{itemize}
\end{mdframed}
It is easily observed that $ \mathbb P_t(\Omega_i, Q_i, t)=\mathop {\lim }\limits_{t_{l+1} - t_l \to 0}   \mathbb P_a(\Omega_i, Q_i, T_l)$. Similarly to FD-EEF, {\it average} FD-EPF or {\it instantaneous} FD-EPF also process a similar frequency-selectiveness property, which can be manifested by direct differentiation or integration. 

\subsection{Frequency-dependent switching logic and  system configuration}
\begin{figure}[htbp]
	\centering
	 \includegraphics[width=0.46\textwidth]{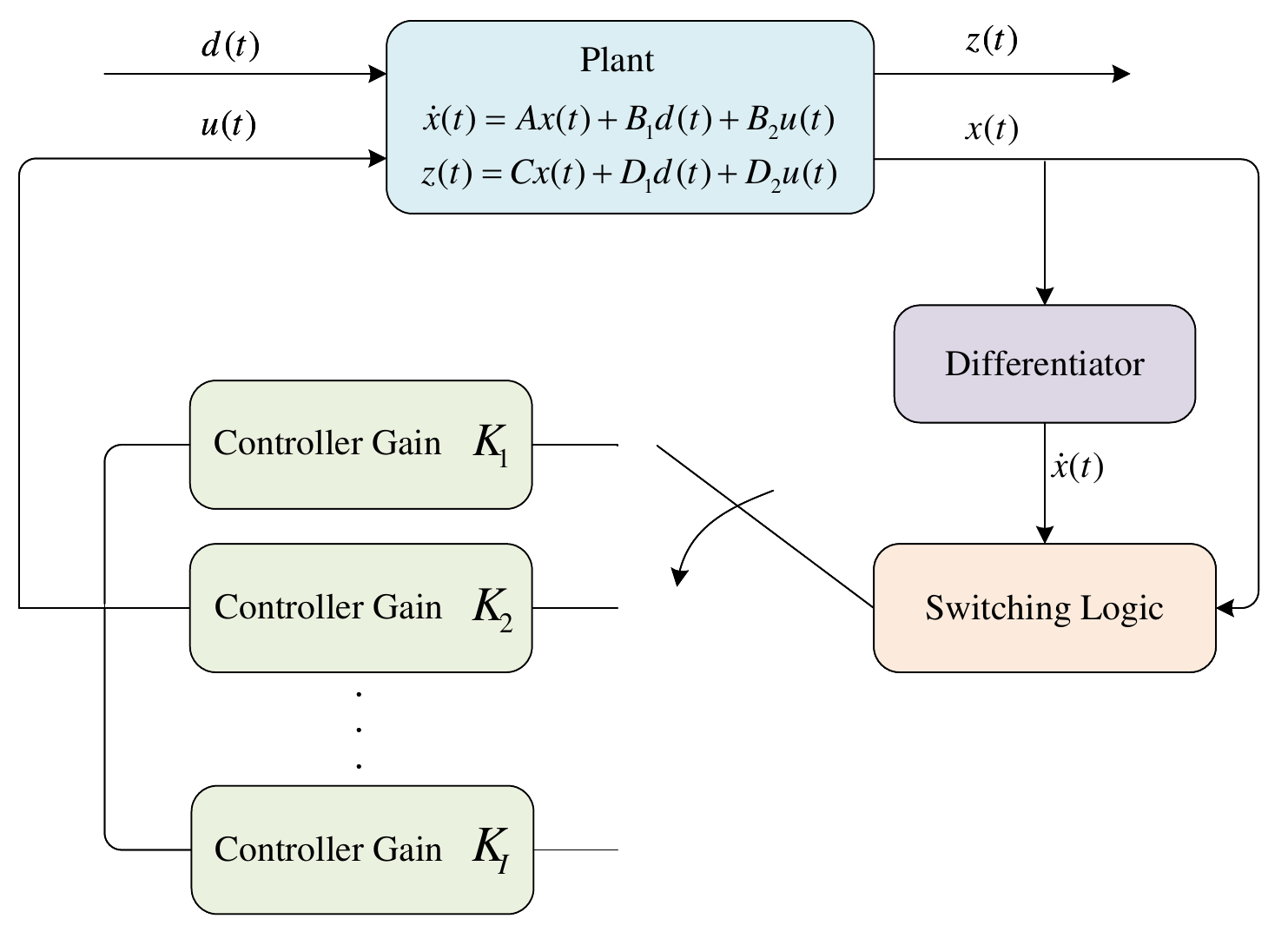}
	\caption{Frequency-dependent switching control system}
	\label{Switching Logic}
\end{figure}

In the light of the frequency selectiveness property of FD-EEF and FD-EPF, picking up the most significant frequency range that dominates the system response in real-time becomes techniquely feasible. In this work, we simplely choose the {\it instantaneous} FD-EPF as the candidate function to construct a switching law. Mathematically, our proposed frequency-dependent switching control (FDSC) scheme can be written as 
\begin{equation}\label{eqn_FDSC}
u(t) = K_{\sigma(t)} x(t),
\end{equation}
where $\sigma(t)$ is the switching signal satisfing the following rule
\begin{equation}\label{switchinglaw}
\sigma(t) = \mathop{{\rm argmax}}\limits_{i\in\{1,2,\ldots,I\}} \mathbb P(\Omega_{i}, Q_{i},t)
=\mathop{{\rm argmax}}\limits_{i\in\{1,2,\ldots,I\}} [\dot x^*(t) \kern 4pt  x^*(t)] (\Psi_{i} \otimes Q_{i}) [ \dot x^*(t) \kern 4pt x^*(t)]^*,
\end{equation}
$\sigma(t)$ is a piece-wise function taking values in $\{1,2,\ldots,I\}$, and $I$ denotes the number of passive controller gains $\{K_1,K_2,\ldots,K_I\}$. The closed-loop paradigm with the FDSC scheme is illustrated in Fig. \ref{Switching Logic}.

Implementing the FDSC scheme, the closed-loop system will be a switched system as follows:
\begin{equation}\label{switchedclosedss}
\begin{split}
\dot x(t) &= \mathbf{A}_{\sigma(t)} x(t) + \mathbf{B}_{\sigma(t)} d(t),\\
z(t) &= \mathbf{C}_{\sigma(t)} x(t) + \mathbf{D}_{\sigma(t)} d(t),
\end{split}
\end{equation}
where
\begin{equation*}
\begin{array}{*{20}{c}}
\mathbf{A}_{\sigma(t)} = A+ B_2 K_{\sigma(t)},  & \mathbf{B}_{\sigma(t)}=B_1, \\
\mathbf{C}_{\sigma(t)} = C+ B_2 K_{\sigma(t)},  & \mathbf{D}_{\sigma(t)}=D_1 .
\end{array}
\end{equation*}
The switching instants are expressed by a sequence:
   \[\mathscr T: \{t_1,t_2,\ldots,t_l,t_{l+1},\ldots\},\]
 and the sequence of switch-in instants of the $i^{th}$ passive control gain is represented by 
\[\mathscr T_i^{i}: \{t_{i_1},t_{i_2},\ldots,t_{i_m},t_{i_{(m+1)}},\ldots\},\] 
while the sequence of its switch-out instants is represented by 
\[\mathscr T_i^{o}: \{t_{i_1+1},t_{i_2+1},\ldots,t_{i_m+1},t_{i_{(m+1)}+1},\ldots\}.\] 
Both $\mathscr T_i^{i}$ and $\mathscr T_i^{o}$ are subsets of $\mathscr T$. The subsystem that is active over the time interval $[t_{l-1}, t_l^-)$ be denoted by $i (i\in \{1,2,\ldots,I\}$), and the index of the subsystem that is active over the time interval $[t_l, t_{l+1}^-)$ be denoted by $j (i\neq j \in \{1,2,\ldots,I\})$. 

Since the passive control gains $\{K_1,K_2,\ldots,K_I\}$ are supposed to be available by applying the gKYP lemma based control design method, the design task is reduced to find out appropriate matrices $Q_i, i=1,2,\ldots,I$, to meet the control objectives. The result given in the following theorem presents conditions that connecting the matrices  $Q_i, i=1,2,\ldots,I$, and the disturbance attenuation performance of the FDSC controller (\ref{eqn_FDSC}).

\begin{mdframed}[hidealllines=true,backgroundcolor=red!2,innerleftmargin=8pt,innerrightmargin=8pt,leftmargin=1pt,rightmargin=1pt]
\begin{theorem}\label{Theorem 1} Suppose $d(t)$ is $L_2$-norm bounded and the initial state is zero, i.e., $x(0)=0$.  Consider the closed-loop switched system \eqref{switchedclosedss}. Assume that there exist a symmetric matrix $P_i$ and positive-definite symmetric matrices $P_{i}^s$, $Q_{i}$ such that the following inequalities hold:
\begin{equation}\label{theorem1-1}
\pmatset{1}{0.1pt}
  \pmatset{0}{0.1pt}
  \pmatset{2}{4pt}
  \pmatset{3}{4pt}
  \pmatset{4}{4pt}
  \pmatset{5}{4pt}
  \pmatset{6}{1pt} 
\begin{pmat}[{.}]
   \mathbf{A}_{i}^* &{\kern 4pt} I\cr
  \end{pmat}(\Phi  \otimes P^s_i+ \Psi_{i}  \otimes  Q_{i}-  \Psi_{j}  \otimes Q_{j} )
\begin{pmat}[{.}]
   \mathbf{A}_{i}^* &{\kern 4pt} I\cr
\end{pmat}^*<0, 
\end{equation}
\begin{equation}\label{theorem1-2}
\begin{array}{l}
 \pmatset{1}{0.1pt}
  \pmatset{0}{0.1pt}
  \pmatset{2}{4pt}
  \pmatset{3}{4pt}
  \pmatset{4}{4pt}
  \pmatset{5}{4pt}
  \pmatset{6}{1pt}
  {\kern 10pt}
\begin{pmat}[{.}]
   \mathbf A_{i}   & {\kern 4pt} \mathbf   B_{i} \cr
   I  &{\kern 4pt} 0 \cr
  \end{pmat}^*
(\Phi  \otimes P_{i} + \Psi_{i}  \otimes  Q_{i}- \Psi_{j}  \otimes  Q_{j} )
 \begin{pmat}[{.}]
   \mathbf A_{i}   & {\kern 4pt} \mathbf   B_i \cr
   I  &{\kern 4pt} 0 \cr
  \end{pmat} +
  \begin{pmat}[{.}]
   \mathbf C_{i}   & {\kern 4pt}\mathbf   D_{i}  \cr
    0   & {\kern 4pt}I \cr
  \end{pmat}^*
  \begin{pmat}[{.}]
   I   & 0  \cr
   0   &-{\gamma_{i}}^2 I\cr
  \end{pmat} \begin{pmat}[{.}]
     \mathbf C_{i}   & {\kern 4pt}\mathbf   D_{i}  \cr
    0   & {\kern 4pt}I \cr
  \end{pmat} <0, 
 \end{array}\end{equation}
for $i,j=1,2,\ldots, I, i \neq j$.  If no sliding motion occurs \cite{Pettersson2005}, then the following statements are true.  \\
(1)  The autonomous closed-loop switched system \eqref{switchedclosedss} with $d(t)=0$ is asymptotically stable under the switching law \eqref{switchinglaw} , i.e., for any given initial condition $x(t_0)$, we have $\lim\limits_{t \to \infty }x(t)\to 0$.\\
(2) The transient disturbance attenuation performance of the closed-loop switched system satisfies
\begin{equation}\label{theorem1-3}
\frac{{{ \int_{T} z^*(t)z(t)dt }}}{{\int_{ T} d^*(t)d(t)dt}}<\max(\gamma_{1}^2,{\gamma_2}^2,\ldots,\gamma_I^2)+\epsilon_{T}
\end{equation}
for any given time interval $ T:=[t_s,t_{m}^-)$, where $t_s$ and $t_m$ are the starting and terminal time instants, respectively, and $\epsilon_{T}$ is the switching caused residue with small value.  \\
(3)  The asymptotic disturbance attenuation performance of the closed-loop switched system satisfies 
\begin{equation}\label{theorem1-4}
\begin{array}{l}
\lim\limits_{t \to \infty } \frac{{{ \int_{0}^{t}z^*(t)z(t)dt }}}{{\int_{0}^{t}d^*(t)d(t)dt}}<\sum\limits_i{ \beta_i\gamma_{i}^2}+\epsilon_\infty,
\end{array}
\end{equation}
where $T_l^i:=[t^i_{l}, (t^i_{l+1})^-)$, $\beta_i= T_l^i/\sum\limits_i T_l^i$, and $\epsilon_\infty$ is the switching caused residue with small value. Moreover, $\beta_i$ increase along with the increase of dominance degree $\alpha(\Omega_i,T_l)$, and vice versa.
\end{theorem}
 \end{mdframed}

\begin{proof}
~\\
\noindent {\textbf {Proof of statement 1)}}\\
Let us define  the following Lyapunov function:
\begin{equation}\label{eqn_T2Proof1}
\mathbb V(t) = x^{*}(t)P^s_{\sigma(t)} x(t).
\end{equation}
Without loss of generality, let us consider an interval $T_l:=[t_l,  t_{l+1}^{-})$ between two consecutive switching instants $t_l$ and $t_{l+1}$ and let $\sigma(t_i)=i$ and $\sigma(t_{i+1}) =j$ and $i, j \in \{1,2,\ldots,I\}$. Then, for $t \in [t_{l}, t_{l+1}^{-})$, subsystem $(\mathbf A_i, \mathbf B_i, \mathbf C_i, \mathbf D_i)$ is activated at the $m^{th}$ time, i.e., $t_l=t_{i_m}$, $t_{l+1}=t_{i_m+1}$. During the time interval from $t_{i_m}$ to $t_{i_m+1}^-$,  the derivative of $\mathbb V (t)$ with respect to time is
\begin{equation}
\dot{{\mathbb V}}(t) = x^*(t)(\mathbf{A}_{i}^* P_i^s+ P_i^s \mathbf{A}_{i} )x(t).
\end{equation}
Multiplying the inequality \eqref{theorem1-1} by $[\dot x^{*}(t)~x^{*}(t)]$ from the left and by its conjugate transpose from the right, we have
\begin{equation}
\begin{array}{l}
x^*(t)(\mathbf{A}_{i}^* P_i^s+ P_i^s \mathbf{A}_{i} )x(t)+[\dot x^*(t) \kern 4pt  x^*(t)] (\Psi_i \otimes Q_i-\Psi_j \otimes Q_j) [ \dot x^*(t) \kern 4pt x^*(t)]^*|_{i\neq j} <0.
\end{array}\end{equation}
From the definition of switching law  \eqref{switchinglaw}, we have
\begin{equation}
\label{lya-der}
\dot{{\mathbb V}}(t) < -[\dot x^*(t) \kern 4pt  x^*(t)] (\Psi_i \otimes Q_i-\Psi_j \otimes Q_j) [ \dot x^*(t) \kern 4pt x^*(t)]^*|_{i\neq j} <0, 
\end{equation}
which implies that closed-loop switched system is asymptotically stable if no sliding motion occurs. 

\noindent {\textbf {Proof of statement 2)}}\\
Multiplying the inequality \eqref{theorem1-2} by $[x^{*}(t)~d^{*}(t)]$ from the left and by its conjugate transpose from the right yields
 \begin{equation}\label{eqn_T1Proof1}
\begin{array}{l}
	{\kern 8pt}
 \pmatset{1}{0.1pt}
\pmatset{0}{0.1pt}
\pmatset{2}{4pt}
\pmatset{3}{4pt}
\pmatset{4}{4pt}
\pmatset{5}{4pt}
\pmatset{6}{1pt} 
\begin{pmat}[{.}]
   \dot{x}^*(t) &   x^*(t) \cr
  \end{pmat}    (\Phi  \otimes P_i+ \Psi_{i}  \otimes  Q_{i}- \Psi_{j}  \otimes  Q_{j}  )
\begin{pmat}[{.}]
   \dot{x}(t) \cr
   x (t)\cr
  \end{pmat}+
 \begin{pmat}[{.}]
   z^{*}(t) &   d^{*}(t)\cr
  \end{pmat} 
   \begin{pmat}[{.}]
   I   & 0  \cr
   0   & -{\gamma_{i}}^2 I\cr
  \end{pmat} 
  \begin{pmat}[{.}]
   z(t) \cr
   d(t) \cr
  \end{pmat} \\
=\frac{d}{dt}(x^{*}(t) P_i x(t))  +\mathbb P_t(\Omega_i, Q_i,t)- \mathbb P_t(\Omega_j, Q_j,t)+z^*(t)z(t)-\gamma_i^2d^*(t)d(t) <0.
\end{array}
\end{equation}
With a slight abuse of notation, let $\bigcup\limits_{l=1}^{m-1}[t_l, t_{l+1}^{-})$ denotes the set including the switching instants $\{t_2,\ldots,t_{m-1}\}$ plus the starting time $t_1:=t_s$ and the terminal time instant $t_m$, then integrating the inequality \eqref{eqn_T1Proof1} over given time interval $T:=[t_s, t_m^{-})$ gives rise to
\begin{equation}\label{eqn_T1Proof2}
\begin{array}{l}
x^{*}(t)P_i x(t)\mid_{t_s}^{t_{m}^-} +\int_{t_s}^{t_{m}^-} (z^*(t)z(t)-\gamma_i^2d^*(t)d(t) )dt 
+\int_{t_s}^{t_{m}^-}  (\mathbb P(\Omega_i, Q_i,t)- \mathbb P(\Omega_j, Q_j,t) )dt < 0.
\end{array}
\end{equation}
\noindent According to the switching law, we have $ \mathbb P(\Omega_{\sigma(t)}, Q_{\sigma(t)},t)-\mathbb P(\Omega_j,Q_j,t) >0, \forall t \in \mathcal T $, thus one can conclude
 \begin{equation}\label{eqn_T1Proof2-conserv}
\begin{array}{l}
\int_{t_s}^{t_{m}^-}(z^*(t)z(t)-\max\{\gamma_1^2, \gamma_2^2,\ldots,\gamma_I^2\}d^*(t)d(t) )dt\\
<\int_{t_s}^{t_{m}^-}(z^*(t)z(t)-\gamma_i^2d^*(t)d(t))dt\\
< -\int_{t_s}^{t_{m}^-}  (\mathbb P(\Omega_i, Q_i,t)- \mathbb P(\Omega_j, Q_j,t) )dt - x^{*}(t)P_i x(t)\mid_{t_s}^{t_{m}^-}.
\end{array}
\end{equation}
Letting 
\begin{equation} 
\begin{array}{l}
\epsilon_T=\max\{0, -\frac{{\int_{t_s}^{t_{m}^-} ( \mathbb P(\Omega_i, Q_i,t)- \mathbb P(\Omega_j, Q_j,t)) dt}}{{\int_{t_s}^{t_{m}^-}d^*(t)d(t)dt}} - \frac{{x^{*}(t)P_i x(t)\mid_{t_s}^{t_{m}^-}}}{{\int_{t_s}^{t_{m}^-}d^*(t)d(t)dt}}\},
\end{array}
\end{equation}
leads us to  
\begin{equation} 
\begin{array}{l}
\frac{\int_{\mathcal T}z^*(t)z(t)dt }{\int_{\mathcal T} d^*(t)d(t)dt} =\frac{  \int_{t_s}^{t_m^-}z^*(t)z(t)dt }{\int_{t_s}^{t_m^-} d^*(t)d(t)dt}<\max(\gamma_{1}^2,{\gamma_2}^2,\ldots,\gamma_I^2)+\epsilon_{\mathcal T}.
\end{array}
\end{equation}

\noindent {\textbf {Proof of statement 3)}}\\
We consider the excited energy of the regulated output over a time interval $T_l^i:=[t^i_{l}, (t^{i}_{l+1})^{-})$, in which the closed-loop subsystem $(\mathbf A_i,\mathbf B_i,\mathbf C_i,\mathbf D_i)$ is activated,  
\begin{equation}
\begin{array}{l}
\int_{t^i_{l}}^{(t^{i}_{l+1})^{-}}z^*(t)z(t)dt \\
<\gamma_{i}^2\int_{t^i_{l}}^{(t^{i}_{l+1})^{-}} d^*(t)d(t)dt - \int_{t^i_l}^{(t^{i}_{l+1})^{-}}  \mathbb (P(\Omega_i, Q_i,t)- \mathbb P(\Omega_j, Q_j,t)) dt- x^{*}(t)P_i x(t)\mid_{t^i_{l}}^{(t^{i}_{l+1})^{-}}.
\end{array}
\end{equation}
The total excited energy of the regulated output is
\begin{equation}
\begin{array}{l}
\lim\limits_{t \to \infty} \int_{0}^t z^*(t)z(t)dt 
=\sum\limits_{i}\sum\limits_{l}\int_{t^i_{l}}^{(t^{i}_{l+1})^{-}}z^*(t)z(t)dt \\
<\sum\limits_{i}\sum\limits_{l}\gamma_{i}^2 \int_{t^i_{l}}^{(t^i_{l+1}){-}} d^*(t)d(t)dt
-\sum\limits_{i}\sum\limits_{l} \int_{t^i_l}^{(t^{i}_{l+1})^{-}}  \mathbb (P(\Omega_i, Q_i,t)- \mathbb P(\Omega_j, Q_j,t)) dt 
-\sum\limits_{i}\sum\limits_{l} x^{*}(t)P_i x(t)\mid_{t_l^i}^{(t^{i}_{l+1})^{-}}.
\end{array}
\end{equation}
Without loss of generality, the input signal can be expressed as $d(t)=\sum\limits_{k}{\mathcal D( \omega_k)e^{\jmath \omega_k t}}$ and its energy can be rewritten in a time-segmented manner as follows:
\[\begin{array}{l}
\sum\limits_i \sum\limits_l  \gamma_{i}^2  \int_{t^i_l}^{(t^{i}_{l+1})^{-}}  d^*(t)d(t)dt\\
=\sum\limits_i \sum\limits_l  \gamma_{i}^2  \int_{t^i_l}^{(t^{i}_{l+1})^{-}}   \sum\limits_{k}\sum\limits_{\kappa} \mathcal D^*(\omega_k) e^{-\jmath\omega_k t}e^{\jmath\omega_{\kappa}t} \mathcal D(\omega_\kappa)dt\\
= \sum\limits_i \sum\limits_l  \gamma_{i}^2  \int_{t^i_l}^{(t^{i}_{l+1})^{-}}  \sum\limits_{k=\kappa}  \mathcal D^*(\omega_k) \mathcal D(\omega_\kappa)e^{\jmath (\omega_{\kappa}-\omega_k)t}  dt 
+ \sum\limits_i \sum\limits_l  \gamma_{i}^2  \int_{t^i_l}^{(t^{i}_{l+1})^{-}}   \sum\limits_{k\neq\kappa} \mathcal D^*(\omega_k) \mathcal D(\omega_\kappa) e^{\jmath (\omega_{\kappa}-\omega_k)t} dt\\
=\mathop {\lim }\limits_{t \to \infty } \sum\limits_i \sum\limits_l  \gamma_{i}^2  \beta_i  \int_{0}^{t}   d^*(t)d(t)dt.
\end{array}\]
Therefore, we have
\begin{equation}\label{theorem1-3-p}
\mathop {\lim }\limits_{t \to \infty } \frac{{{ \int_{0}^{t}z^*(t)z(t)dt }}}{{\int_{0}^{t}d^*(t)d(t)dt}}< \sum_i \beta_i\gamma_{i}^2+\epsilon_\infty,
\end{equation}
\noindent where 
\begin{equation} 
\begin{array}{l}
\epsilon_\infty=\max\{0,  -\frac{{\lim\limits_{t \to \infty } \int_{0}^{t}  (\mathbb P(\Omega_i, Q_i,t)- \mathbb P(\Omega_j, Q_j,t)) dt}}{\lim\limits_{t \to \infty}{\int_{0}^{t}d^*(t)d(t)dt}}- \frac{{\sum\limits_{i}\sum\limits_{l} x^{*}(t)P_i x(t)\mid_{t_l^i}^{(t^{i}_{l+1})^{-}}}}{\lim\limits_{t \to \infty}{\int_{0}^{t}d^*(t)d(t)dt}}\}.
\end{array}
\end{equation}
Now, let us focus on the state response of the closed-loop system during a time slot: $ T_l:=[t_{l}, t_{l+1}^-)$, where $t_{l}, t_{{l+1}}$ are two  consecutive switching time instants. The system state response and its derivative over  $[t_{l}, t_{l+1}^-)$ are given as follows:
\begin{equation}\label{eqn_x_state_response}
\begin{array}{l}
x(t)=\sum\limits_{k}(\mathcal D( \omega_k)e^{\jmath \omega_k t}I-\mathcal D( \omega_k)e^{\jmath\omega_{k}t_{l}}e^{\mathbf A_i (t-t_{l})})(\jmath \omega_k I-\mathbf A_i)^{-1}\mathbf B_i+e^{\mathbf A_i(t-t_{l})}x(t_{l}),
\end{array}
\end{equation}
\begin{equation}
\begin{array}{l}
\dot x(t)=\sum_{k}(\jmath\omega_k \mathcal D(\omega_k)  e^{\jmath \omega_k t}I-\mathcal D( \omega_k) \mathbf A_i e^{\jmath\omega_{k}t_{l}}e^{\mathbf A_i (t-t_{l})})(\jmath \omega_k I-\mathbf A_i)^{-1}\mathbf B_i+\mathbf A_i e^{A_i (t-t_{l})}x(t_{l}) ,
\end{array}
\end{equation}
where $x(t_{l})$ is the system state at the switching instant. A detailed analysis of the FD-EPFs can be presented as follows:
\[\begin{array}{l}
\mathbb P(\Omega_i, Q_i, t)
=  [\dot x^*(t) \kern 4pt  x^*(t)] (\Psi \otimes Q_i) [ \dot x^*(t) \kern 4pt x^*(t)]^*\\
={\mathbb P_a}(\Psi_i, Q_i, \mathbf A_i, \mathbf B_i, t)  +{\mathbb P_e}(\Psi_i, Q_i, \mathbf A_i, \mathbf B_i, e^{\mathbf A_i (t-t_{l})} ),
\end{array}\]
where the first term is 
\[\begin{array}{l}
{\mathbb P_a}(\Psi_i, Q_i, \mathbf A_i, \mathbf B_i,t ) \\
	=\Gamma (\Psi_i, Q_i, \sum\limits_{\omega_k} e^{\jmath \omega_k t}(\jmath \omega_k I-\mathbf A_i)^{-1}\mathbf B_i,   \sum\limits_{\omega_k} \jmath \omega_k e^{\jmath \omega_k t}(\jmath \omega_k I-\mathbf A_i)^{-1}\mathbf B_i) \\
	= \sum\limits_{\omega_k, \omega_\kappa, k=\kappa}\tau(\Psi_i, \omega_k, \omega_\kappa) e^{-\jmath\omega_k t }e^{\jmath\omega_\kappa t }[(\jmath \omega_\kappa I-\mathbf A_i)^{-1}\mathbf B_i]^* Q_i [(\jmath \omega_k I-\mathbf A_i)^{-1}\mathbf B_i)] \\
{\kern 10pt}+ \sum\limits_{\omega_k, \omega_\kappa, k\neq\kappa}\tau(\Psi_i, \omega_k, \omega_\kappa) e^{-\jmath\omega_k t }e^{\jmath\omega_\kappa t }[(\jmath \omega_\kappa I-\mathbf A_i)^{-1}\mathbf B_i]^* Q_i [(\jmath \omega_k I-\mathbf A_i)^{-1}\mathbf B_i)] \\
	= \underbrace{\sum\limits_{\begin{subarray}{c} \omega_k, \omega_\kappa, k=\kappa\\ \omega_k\in \Omega_i \end{subarray}}
		\tau(\Psi_i, \omega_k, \omega_k) [(\jmath \omega_k I-\mathbf A_i)^{-1}\mathbf B_i]^* Q_i [(\jmath \omega_k I-\mathbf A_i)^{-1}\mathbf B_i)]} _{\mathbb P_a^{in}(\Psi_i, Q_i, \mathbf A_i, \mathbf B_i)>0} \\
{\kern 10pt}+  \underbrace{\sum\limits_{\begin{subarray}{c} \omega_k, \omega_\kappa, k=\kappa\\ \omega_k\notin \Omega_i \end{subarray}}
		\tau(\Psi_i, \omega_k, \omega_k)  [(\jmath \omega_k I-\mathbf A_i)^{-1}\mathbf B_i]^*Q_i[(\jmath \omega_k I-\mathbf A_i)^{-1}\mathbf B_i)]} _{\mathbb P_a^{out}(\Psi_i, Q_i, \mathbf A_i, \mathbf B_i)<0} \\ 
{\kern 10pt}+  \underbrace{\sum\limits_{\omega_k, \omega_\kappa, k\neq\kappa}\tau(\Psi_i, \omega_k, \omega_\kappa) e^{\jmath(\omega_\kappa-\omega_k) t } [(\jmath \omega_\kappa I-\mathbf A_i)^{-1}\mathbf B_i]^*Q_i[(\jmath \omega_k I-\mathbf A_i)^{-1}\mathbf B_i)]} _{\mathbb P_e(\Psi_i, Q_i, \mathbf A_i, \mathbf B_i, e^{\jmath\omega_k t})} \\ 
\end{array}\]
with 
\[\tau(\Psi_i,\omega_k,\omega_\kappa)= \left\{ \begin{array}{l}
(\varpi_l-\omega_k)(\omega_\kappa+\varpi_l) ,\; ~{\rm if}~ \Omega_i  \in \Omega_l \\
(\varpi_2-\omega_k)(\omega_\kappa-\varpi_1),\; {\rm if} ~\Omega_i  \in \Omega_m \\
(\omega_k-\varpi_h)(\omega_\kappa+\varpi_h),\; {\rm if} ~\Omega_i  \in \Omega_h,\\
\end{array} \right.\]
and the second term can be written as
\[ \begin{array}{l}
	{\mathbb P_e}(\Psi_i, Q_i, \mathbf A_i, \mathbf B_i, e^{\mathbf A_i (t-t_l)}) \\
	= \sum\limits_{\omega_k}\Gamma (\Psi_i, Q_i, \sum\limits_{\omega_k} e^{\jmath \omega_k t}(\jmath \omega_k I-\mathbf A_i)^{-1}\mathbf B_i,  \jmath\omega_k e^{\jmath \omega_k t} (\jmath \omega_k I-\mathbf A_i)^{-1}\mathbf B_i)  \\
	{\kern 12pt}+\sum\limits_{\omega_k}\Gamma (\Psi_i, Q_i, \sum\limits_{\omega_k} e^{\jmath \omega_k t}(\jmath \omega_k I-\mathbf A_i)^{-1}\mathbf B_i,   \mathbf A_i e^{\jmath\omega_{k}t_l}e^{\mathbf A_i (t-t_l)}(\jmath \omega_k I-\mathbf A_i)^{-1}\mathbf B_i)  \\
	{\kern 12pt}+ \sum\limits_{\omega_k} \Gamma (\Psi_i, Q_i, e^{\jmath \omega_k t}(\jmath \omega_k I-\mathbf A_i)^{-1}\mathbf B_i, \mathbf A_ie^{\mathbf A_i (t-t_l)}x(t_l))  \\
	{\kern 12pt}+  \sum\limits_{\omega_k}\Gamma (\Psi_i, Q_i,\sum\limits_{\omega_k} e^{\jmath\omega_{k}t_l}e^{\mathbf A_i (t-t_l)}(\jmath \omega_k I-\mathbf A_i)^{-1}\mathbf B_i, \jmath\omega_k e^{\jmath \omega_k t} (\jmath \omega_k I-\mathbf A_i)^{-1}\mathbf B_i)  \\
	{\kern 12pt}+  \sum\limits_{\omega_k}\Gamma (\Psi_i, Q_i, \sum\limits_{\omega_k} e^{\jmath\omega_{k}t_l}e^{\mathbf A_i (t-t_l)}(\jmath \omega_k I-\mathbf A_i)^{-1}\mathbf B_i, \mathbf A_i e^{\jmath\omega_{k}t_l}e^{\mathbf A_i (t-t_l)}(\jmath \omega_k I-\mathbf A_i)^{-1}\mathbf B_i)  \\
	{\kern 12pt}+\sum\limits_{\omega_k}  \Gamma (\Psi_i, Q_i, e^{\jmath\omega_{k}t_l}e^{\mathbf A_i (t-t_l)}(\jmath \omega_k I-\mathbf A_i)^{-1}\mathbf B_i, \mathbf A_ie^{\mathbf A_i (t-t_l)}x(t_l))  \\
	{\kern 12pt}+ \sum\limits_{\omega_k} \Gamma (\Psi_i,  Q_i,e^{\mathbf A_i (t-t_l)}x(t_l),  \jmath\omega_k e^{\jmath \omega_k t} (\jmath \omega_k I-\mathbf A_i)^{-1}\mathbf B_i)  \\
	{\kern 12pt}+ \sum\limits_{\omega_k} \Gamma (\Psi_i, Q_i, e^{\mathbf A_i (t-t_l)}x(t_l),  \mathbf A_i e^{\jmath\omega_{k}t_l}e^{\mathbf A_i (t-t_l)}(\jmath \omega_k I-\mathbf A_i)^{-1}\mathbf B_i)  \\
	{\kern 12pt}+ \Gamma (\Psi_i, Q_i, e^{\mathbf A_i (t-t_l)}x(t_l), \mathbf A_ie^{\mathbf A_i (t-t_l)}x(t_l))  \\
\end{array}\]
 with
\[
\Gamma (\Psi_i, Q_i, X,Y)= \sum\limits_{\omega_k, \omega_\kappa} \tau(\Psi_i,\omega_k,\omega_\kappa)XQ_iY.
\]
As the closed-loop switched system is stable,  the integral of the terms with $e^{\jmath(\omega_\kappa-\omega_k) t}$ will be uniformly bounded, i.e.,   
\[\mathop {\lim }\limits_{t \to \infty } \int_0^ t \mathbb P_e(\Psi_i, Q_i, \mathbf A_i, \mathbf B_i, e^{\jmath\omega_k t})  dt < \mathbb B_{1i},\]
where $\mathbb B_{1i}$ denotes the $i$-th upper bound with finite value.  Since $\mathbf A_i$ are Hurwitz matrices and the closed-loop system is asymptotically stable, it can be proved that all terms associated with $e^{\mathbf A_i (t-t_l)}$ are uniformly bounded according to a simple extension of the Lemma A.1. in reference \cite{fdscIwasaki2005time}. Furthermore, we have
\[\mathop {\lim }\limits_{t \to \infty } \int_0^ t \mathbb P_e(\Psi_i, Q_i, \mathbf A_i, \mathbf B_i,  e^{\mathbf A_i (t-t_l)})dt  < \mathbb B_{2i},\]
where $\mathbb B_{2i}$ also represents a finite upper bound. 

With the above analysis on FD-EPFs as well as the relationship between FD-EEFs and FD-EPFs, we have
\[\begin{array}{l}
\mathop {\lim }\limits_{t \to \infty } \mathbb S(\Omega_i, Q_i, [0,t))\\
=\mathop {\lim }\limits_{t \to \infty } \int_0^t  \mathbb P_t(\Omega_i, Q_i, t) dt \\
= \mathop {\lim }\limits_{t \to \infty } \int_0^t  \mathbb P_a^{in} (\Omega_i, Q_i, \mathbf A_i, \mathbf B_i) dt +\mathop {\lim }\limits_{t \to \infty } \int_0^t  \mathbb P_a^{out} (\Omega_i, Q_i, \mathbf A_i, \mathbf B_i)  dt \\
{\kern 11pt}+\mathop {\lim }\limits_{t\to \infty } \int_0^t \mathbb P_e(\Psi_i, Q_i, \mathbf A_i, \mathbf B_i, e^{\jmath\omega_k t})  dt 
+ \mathop {\lim }\limits_{t \to \infty } \int_0^t \mathbb P_e(\Psi_i, Q_i, \mathbf A_i, \mathbf B_i,  e^{\mathbf A_i (t-t_\kappa)})dt  \\
< \mathop {\lim }\limits_{t \to \infty } (\mathbb P_a^{in} (\Omega_i, Q_i, \mathbf A_i, \mathbf B_i) +\mathbb P_a^{out} (\Omega_i, Q_i, \mathbf A_i, \mathbf B_i)) t +\mathbb B_{1i}+ \mathbb B_{2i}.
\end{array}\]

\noindent  It can be observed that the value of the FD-EPFs mainly depends on the {\it invariant} in-band component $ \mathbb P_a^{in} (\Omega_i, Q_i, \mathbf A_i, \mathbf B_i) $ and the  {\it invariant} out-of-band component $\mathbb P_a^{out} (\Omega_i, Q_i, \mathbf A_i, \mathbf B_i)$. On the other hand, the {\it invariant} components coincide with the {\it average}  FD-EPF of its associated subsystem over infinite time horizon, specifically, we have 
\[\begin{array}{l}
\lim\limits_{t\to \infty } \mathbb P_a (\Omega_i, Q_i, t)= \lim\limits_{t\to \infty } (\mathbb P_a^{in} (\Omega_i, Q_i, \mathbf A_i, \mathbf B_i)  + \mathbb P_a^{out} (\Omega_i, Q_i, \mathbf A_i, \mathbf B_i)).
\end{array}\]
According to our switching mechanism, it is more likely to activate the subsystem which is associated to larger {\it average} FD-EPF over a long time horizon.  The changing  principal of a group of terms with respect to the variation of $\alpha_i$ can be derived and presented in the following Table \ref{table}. 

\begin{table}[h]
{\begin{center}\label{relationship}
\tabcolsep=1cm
\renewcommand\arraystretch{1.5}
            \makeatletter\def\@captype{table}\makeatother\caption{ Increase and decrease of a group of terms with changing $\alpha_i$}
 \begin{tabular}{|c|c|c|}
\hline
  &$\alpha_i(\Omega_{i}, T_l) \uparrow$                            & $\alpha_i(\Omega_{i}, T_l) \downarrow$    \\
 \hline
 $\left|{P_a^{in} (\Omega_i, Q_i, \mathbf A_i, \mathbf B_i)}\right|$  &  $\uparrow $  &   $\downarrow $      \\
\hdashline
$\left|{P_a^{out} (\Omega_i, Q_i, \mathbf A_i, \mathbf B_i)}\right|$ &  $\downarrow $  &   $\uparrow $      \\
 \hline
${P_a (\Omega_i, Q_i, \mathbf A_i, \mathbf B_i)}$ &  $\uparrow $  &   $\downarrow $      \\
\hdashline
$\max\{{P_a (\Omega_j, Q_j, \mathbf A_j, \mathbf B_j)}\}, j\neq i$ &  $\downarrow $  &   $\uparrow $      \\
\hline
$\mathcal T_i =\sum_{l} T_{l}^i$    &   $\uparrow $     & $\downarrow $ \\
\hdashline
$\sum_j \mathcal T_j =\sum_j \sum_{l} T_{l}^j , j\neq i$    &   $\downarrow $     &  $\uparrow $   \\
\hline
$\beta_i$    &   $\uparrow $     &   $\downarrow $    \\
\hline
\end{tabular}\label{table}\end{center}}
\end{table}
\noindent This concludes the proof of our theorem. 
\end{proof}

\begin{remark} {\it
Besides the matrices $Q_i, i=1,2,\ldots,I$, to be designed, the achievable performance of our FDSC controller also depends on the passive control gains $K_i, i=1,2,\ldots,I$. With the assumption that the disturbance is frequency dominated with respect to $\Omega_i$ , the in-band passive controller always has to be included (i.e., setting $K_i=K^{in}$). In case that there does not exist {\it a priori} time-frequency characteristics about the out-of-band components, picking up only one controller gain from the available out-of-band controller pool $K^{out}_i, i=1,2,\ldots,I$, becomes the simplest option. To choose the most appropriate out-of-band controller, a promising way is to compare the maximum singular value gap between the transfer functions resulting from the in-band and out-of-band. The gap function can be explicitly rewritten as:    
  \begin{equation} 
 gap(K^{in}, K^{out}_i, \omega)=\sigma_{max}( G(\jmath\omega), K^{in})-\sigma_{max} (G(\jmath\omega), K^{out}_i), ~ \forall \omega\in\Omega_i,~ i=1,2,\ldots,I.
       \end{equation}
Since the proposed FDSC scheme is aimed to balance the in-band and the out-of-band attenuation performance in a sophisticated way, $K^{out}_i$ will be a good choice if the value of the gap function ${\rm gap}(K^{in}, K^{out}_i, \omega)$ are positive over the entire frequency band $\Omega^{in}$ (or its major subsets) and negative over the entire out-of-band  $\Omega^{out}$ (or its major subsets). In case that the out-of-band components are dominated over multiple discriminative frequency bands, introducing multiple associated out-of-band controllers into the FDSC scheme becomes an attemptable option.}
\end{remark}

\begin{remark} 
{\it To avoid unacceptable out-of-band performance both in finite or infinite time horizon, a simple but reasonable way is to set $\gamma_j=\gamma_{tol}, j\neq i$, in the LMI condition  (\ref{theorem1-2}), then produce the $Q_i$ by solving an optimization problem as follows:}
\begin{equation}\label{opt_problem_2}
\begin{array}{*{20}{c}}
\mathop {\min }\limits_{\gamma_{i},P_i^{s},P_{i},Q_{i}}  \gamma_{i},\\
{\rm {subject \;\; to\;\;} }  (\ref{theorem1-1}) (\ref{theorem1-2})  .
\end{array}
\end{equation}
\end{remark}

\begin{remark} {\it
In the gKYP lemma based passive controller design \cite{fdsciwasaki2007feedback}, it is known that there exists a gap between the {\it a priori} upper bound $\gamma$ and the  {\it a posteriori} actual in-band disturbance  attenuation performance, which can be accurately evaluated by frequency sweeping maximum singular value $\sigma_{max}(G(\jmath\omega), K), \omega \in \Omega$. Noting that the design conditions in Theorem \ref{Theorem 1} are also sufficient, thus the indices $\gamma_i, i=1,2,\ldots,I$, as well as the {\it a priori} bounds are larger than the actual disturbance attenuation performance,  the existence of the gap can be observed from the inequality \eqref{eqn_T1Proof2-conserv}. Unfortunately, it is hard to give accurate {\it a posteriori}  disturbance attenuation performance analysis directly from the  maximum singular value of subsystems $\sigma_{max}(G(\jmath\omega), K_i), \omega \in \Omega_i, i=1,2,\ldots,I$, as accurate frequency-domain analysis of switched systems is intrinsically difficult. However, maximum singular values are also useful to characterize the attenuation performance of FDSC under certain circumstances. Specifically,  in case that the frequency dominant with respect to $\Omega_i$ is very high, e.g. close to $\alpha_i\rightarrow 1$ or even be strictly frequency-limited $\alpha_i\rightarrow 1$ over a time window $T$,  the control gain $K_i$ will be the dominate one among all passive gains, in other word, the behavior of FDSC will mimic the in-band passive controller, thus we have   
       \begin{equation}
    \frac{{{ \int_{T}z^*(t)z(t)dt }}}{{\int_{ T}d^*(t)d(t)dt}}<\sigma_{max}(G(\jmath\omega), K_i)+\epsilon< \gamma_i+\epsilon, \omega \in \Omega_i,
       \end{equation}  
where $\epsilon$ is a small positive constant.  }
\end{remark}

\begin{remark} {\it
In our proposed FDSC scheme, only the FD-EPF is facilitated to formulate the switching law using the state-derivative. Noticing that both FD-EPF and FD-EEF possess frequency-selectiveness properties, it is probably feasible to develop a FD-EEF or FD-EEF $\&$ FD-EPF based switching law as alternatives, in which the state integral will be involved. Although further discussion on this issue is beyond the scope of this paper, to some extent, the design methodology behind our switching law coincides with the underlying principal of the classical Proportional-Integral-Derivative controller (PID) control theory.}
\end{remark}

\begin{remark} {\it
Theorem \ref{Theorem 1} only confirms the stability and disturbance attenuation performance of the closed-loop switched system under the sliding motion free assumption. In fact, sliding motion very rarely occurs in practical applications due to the limited precision and time-varying disturbance. However, providing a theoretical analysis on the existence of sliding motion \cite{berrada2018sliding, Pettersson2005, Fan2020} under our proposed switching law is still meaningful to be discussed in future work. Moreover, it is known that the sliding motion can be strictly avoided by slightly modifying the switching law, for example, introducing hysteresis or minimum dwell time between switching instants \cite{UShaked2012}.}
\end{remark}

\section{Simulation}

The code and the raw data of the presented numerical results are available as noted in Figure \ref{fig:linkcodedata}.

\begin{figure}[h]
  \caption{Code and Data Availability.}\label{fig:linkcodedata}
  \begin{center}
    \fbox{
      \begin{minipage}{.9\linewidth}
        The source code of the implementations used to compute the presented results is available from
        \begin{center}
          \href{https://doi.org/10.5281/zenodo.7990805}{\texttt{doi:10.5281/zenodo.7990805}}
        \end{center}
        under the CC-BY SA license and is authored by Jingjing Zhang.
      \end{minipage}
    }
  \end{center}
\end{figure}

Consider the longitudinal axis flight control system design of an aircraft with
the decoupled linearized longitudinal dynamical equation of motion described as
follows, see, also \cite{adams2012robust} for a detailed description of this
common benchmark example:
\begin{equation*}% \label{aircraft model}
	\begin{array}{l}
		 \pmatset{1}{0.1pt}
		\pmatset{0}{0.1pt}
		\pmatset{2}{4pt}
		\pmatset{3}{4pt}
		\pmatset{4}{4pt}
		\pmatset{5}{4pt}
		\pmatset{6}{1pt} 
		\begin{pmat}[{.}]
			\dot \alpha(t) \cr
			\dot q(t)\cr
		\end{pmat}= A_{long} \begin{pmat}[{.}]
			 \alpha(t) \cr
			 q(t)\cr
		\end{pmat}+ B_{long}^u \begin{pmat}[{.}]
			\delta_E(t)\cr
			\delta_{PTV}(t)\cr
		\end{pmat} + B_{long}^w d(t) ,\\
	{\kern 18pt}	z(t) =C_{long}	
		 \pmatset{1}{0.1pt}
		\pmatset{0}{0.1pt}
		\pmatset{2}{4pt}
		\pmatset{3}{4pt}
		\pmatset{4}{4pt}
		\pmatset{5}{4pt}
		\pmatset{6}{1pt} 
		\begin{pmat}[{.}]
			\alpha(t)\cr
			q(t)\cr
		\end{pmat} + D_{long}\begin{pmat}[{.}]
			\delta_E(t)\cr
			\delta_{PTV}(t)\cr
		\end{pmat},
	\end{array}
\end{equation*}
where $\alpha(t)$ is the  angle of attack , $q(t) $ is the perturbational  pitch rate, $ \delta_E(t)$ is the symmetric horizontal tail deflection, $\delta_{PTV}(t)$ is the  symmetric pitch thrust vectoring nozzle deflection. $d(t)$ is the disturbance caused by wind shear or  mechanical  oscillation, $z(t)$ is the regulated output that balances the state variation and control input. 

We borrow the following data from \cite{yang2010reliable}:
\[\begin{array}{l}
	A_{long}=	 \pmatset{1}{0.1pt}
	\pmatset{0}{0.1pt}
	\pmatset{2}{4pt}
	\pmatset{3}{4pt}
	\pmatset{4}{4pt}
	\pmatset{5}{4pt}
	\pmatset{6}{1pt} 
	\begin{pmat}[{.}]
		-1.175 &{\kern 2pt}0.9871\cr
		-8.458 &{\kern 2pt}-0.8776\cr
	\end{pmat},~~
	B^u_{long}=	\begin{pmat}[{.}]
		-0.194 &{\kern 2pt}-0.03593\cr
		-19.29 &{\kern 2pt}-3.803\cr
	\end{pmat},\\
	B^w_{long}=\pmatset{1}{0.1pt}
	\pmatset{0}{0.1pt}
	\pmatset{2}{4pt}
	\pmatset{3}{4pt}
	\pmatset{4}{4pt}
	\pmatset{5}{4pt}
	\pmatset{6}{1pt} 
		\begin{pmat}[{.}]
		1\cr
		4\cr
	\end{pmat},~~C_{long}=	\begin{pmat}[{.}]
		0 &{\kern 2pt}4\cr
		0 &{\kern 2pt}0\cr
		0 &{\kern 2pt}0 \cr
	\end{pmat},~~
	D_{long}=	\begin{pmat}[{.}]
		0 &{\kern 2pt}0\cr
		2 &{\kern 2pt}0\cr
		0 &{\kern 2pt}2\cr
	\end{pmat}.
\end{array}\]

Hereby we use the concept of LF/MF/HF and consider the following three finite-frequency ranges by simply letting $10^0, 10^1$ as the crossover frequencies, i.e.
\begin{equation}\label{configuration}
	\begin{split}
		{\rm LF } {\kern 12pt} &  \omega \in  \Omega_{1}: (-1, 1), \\
		{\rm MF}{\kern 12pt}   &\omega \in   \Omega_{2}:  (-10, -1] \cup [1, 10), \\
		{\rm HF }{\kern 12pt}   &\omega \in  \Omega_{3}: (-\infty, -10] \cup [10, +\infty).
	\end{split}
\end{equation}

Suppose the dominating frequency components are located in the HF range $\Omega_3$. By applying the gKYP lemma based finite-frequency control design approach established by Iwasaki and Hara,  a group of passive state-feedback controllers can be derived. The control gains as well as their in-band disturbance attenuation performance are given for the LF case, MF case, HF case, respectively,  as follows: 
\[\begin{array}{l}
	K_{f_1} =\pmatset{1}{0.1pt}
	\pmatset{0}{0.1pt}
	\pmatset{2}{4pt}
	\pmatset{3}{4pt}
	\pmatset{4}{4pt}
	\pmatset{5}{4pt}
	\pmatset{6}{1pt} 
	\begin{pmat}[{.}]
		0.1048  &{\kern 2pt} 15.0897\cr
		0.0205   & {\kern 2pt}2.9760\cr
	\end{pmat}, {\kern 21pt} \sigma_{max}(G(\jmath\omega), K_{f_1}) <0.3443,{\kern 2pt} \forall \omega \in \Omega_1,\\
	K_{f_2} = \pmatset{1}{0.1pt}
	\pmatset{0}{0.1pt}
	\pmatset{2}{4pt}
	\pmatset{3}{4pt}
	\pmatset{4}{4pt}
	\pmatset{5}{4pt}
	\pmatset{6}{1pt} 
	\begin{pmat}[{.}]
		-1.6368  &{\kern 2pt} 39.1121\cr
		-0.3229   & {\kern 2pt}7.7111\cr
	\end{pmat}, {\kern 12pt} \sigma_{max}(G(\jmath\omega),K_{f_2}) <0.4025, {\kern 2pt}\forall \omega \in \Omega_2,\\
	K_{f_3} =  \pmatset{1}{0.1pt}
	\pmatset{0}{0.1pt}
	\pmatset{2}{4pt}
	\pmatset{3}{4pt}
	\pmatset{4}{4pt}
	\pmatset{5}{4pt}
	\pmatset{6}{1pt} 
	\begin{pmat}[{.}]
		-0.4217 & {\kern 2pt}-0.0433\cr
		-0.0832   &{\kern 2pt}-0.0085\cr	
	\end{pmat},{\kern 5pt} \sigma_{max}(G(\jmath\omega),K_{f_3}) <0.1714, {\kern 2pt}\forall \omega \in \Omega_3.
\end{array}\]
For comparison, the standard passive state-feedback controller resulting optimal disturbance attenuation performance over EF, i.e., the standard optimal $H_\infty$ controller as well as its EF disturbance attenuation performance are generated as follows:
\[\begin{array}{l}
	K_{e} = \pmatset{1}{0.1pt}
	\pmatset{0}{0.1pt}
	\pmatset{2}{4pt}
	\pmatset{3}{4pt}
	\pmatset{4}{4pt}
	\pmatset{5}{4pt}
	\pmatset{6}{1pt} 
	\begin{pmat}[{.}]
		-1.6360 & {\kern 2pt} 39.0849 \cr
		-0.3228    &{\kern 2pt} 7.7057\cr
	\end{pmat}, {\kern 10pt}\sigma_{max}(G(\jmath\omega),K_e) <0.7125 , {\kern 2pt}\forall \omega \in \Omega_e:(-\infty,+\infty).
\end{array}\]

\begin{figure}[!htbp]
	\centering
	\includegraphics[width=0.9\textwidth]{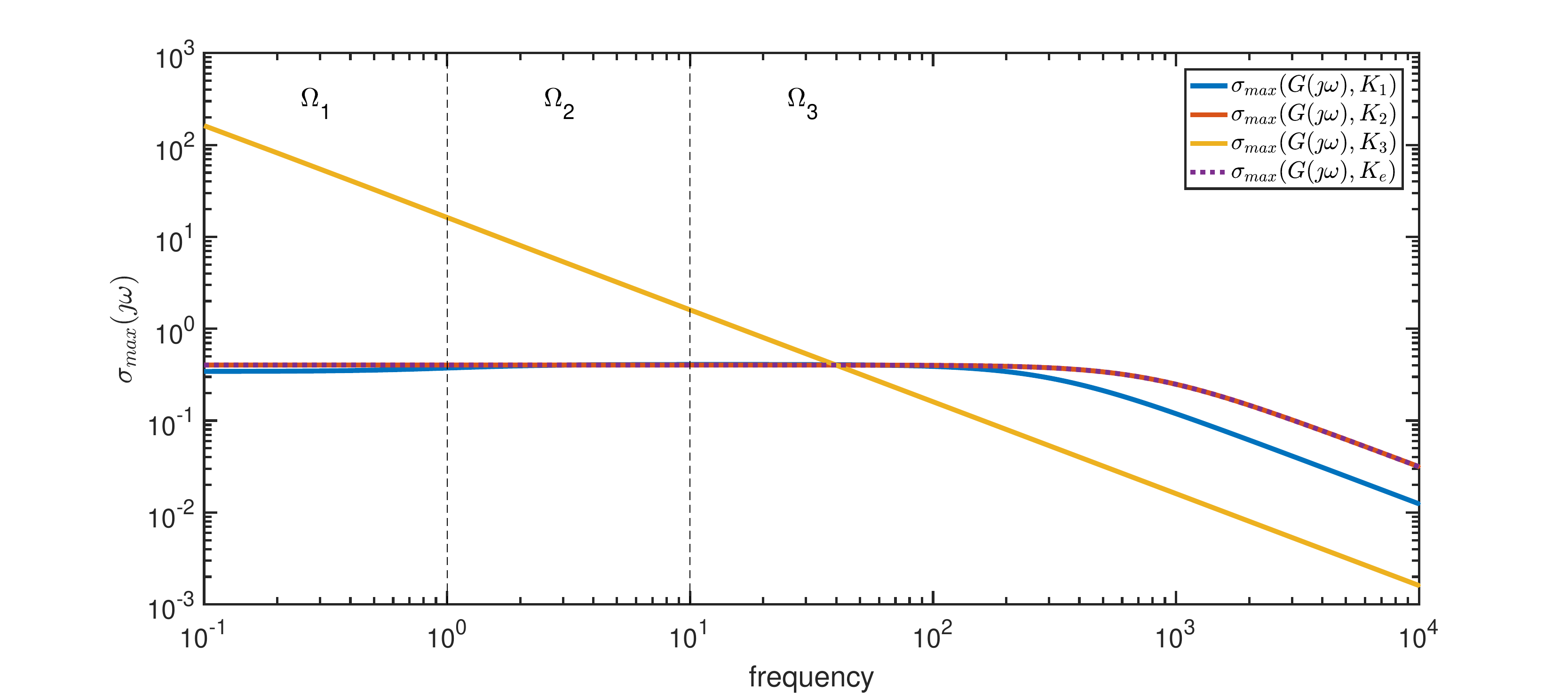}
	\caption{The maximum singular value of closed-loop system model with passive controllers}
	\label{maximum singular value}
\end{figure}

To show the disturbance attenuation performance over different frequency ranges, the maximum singular value of closed-loop system models with passive controllers $K_{f_1}, K_{f_2}, K_{f_3}, K_e$ are plotted, see Figure \ref{maximum singular value}.  Obviously, $K_{f_3}$ provides best HF disturbance attenuation performance while it suffers from a significantly deteriorating LF performance.    

\begin{figure}[t!]
    \centering
    \begin{subfigure}[b]{0.9\textwidth}
           \centering
           \includegraphics[width=\textwidth]{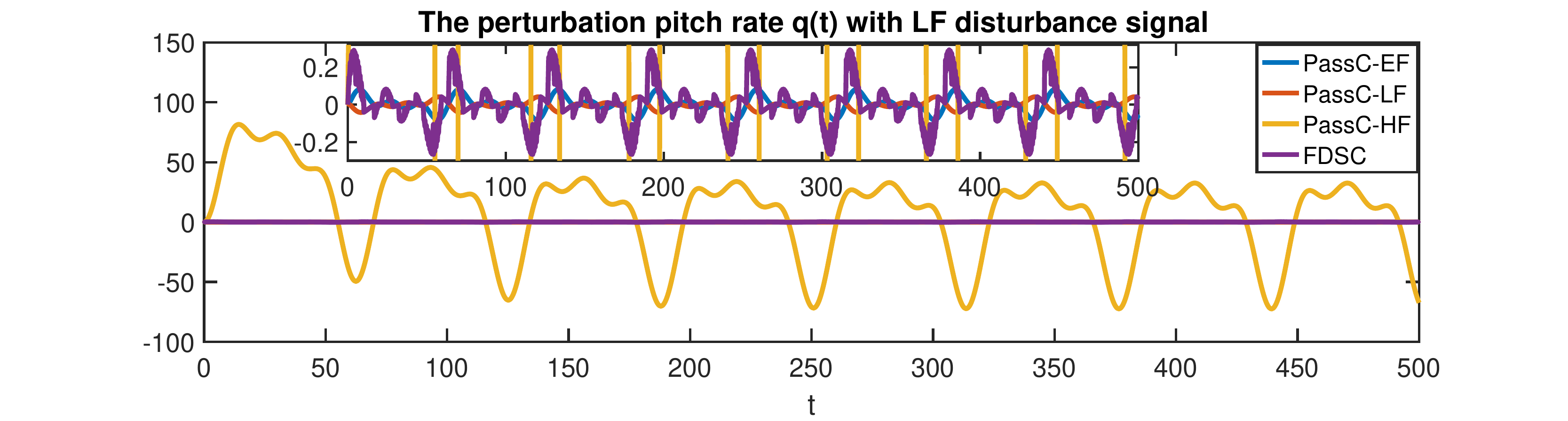}
            \caption{}
             \label{q-LF}
    \end{subfigure}
    \begin{subfigure}[b]{0.9\textwidth}
            \centering
            \includegraphics[width=\textwidth]{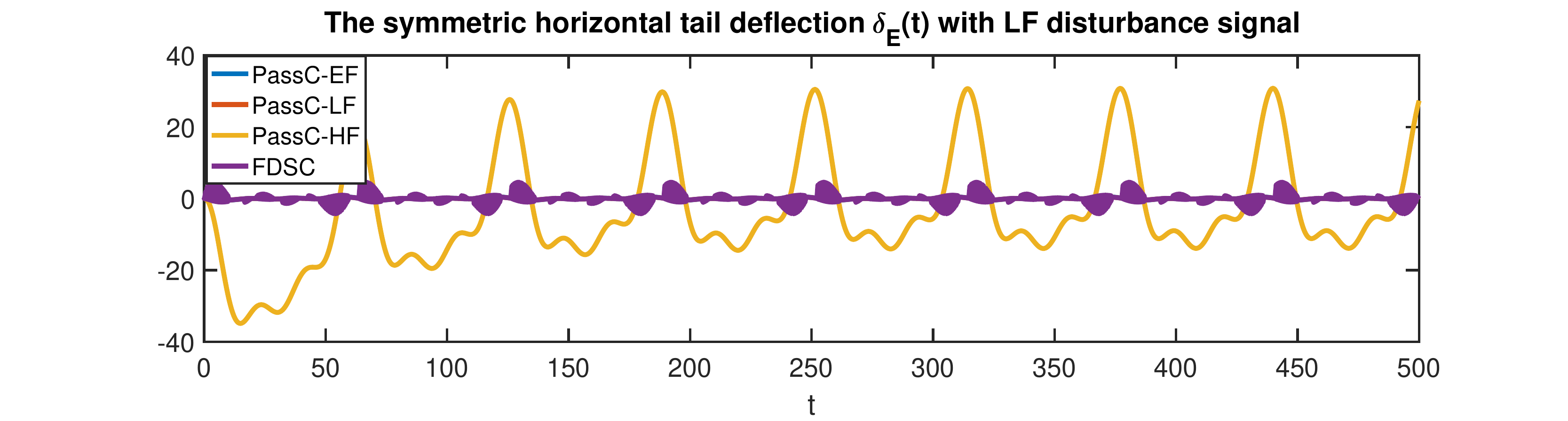}
            \caption{}
             \label{E-LF}
    \end{subfigure}
    \begin{subfigure}[b]{0.9\textwidth}
            \centering
            \includegraphics[width=\textwidth]{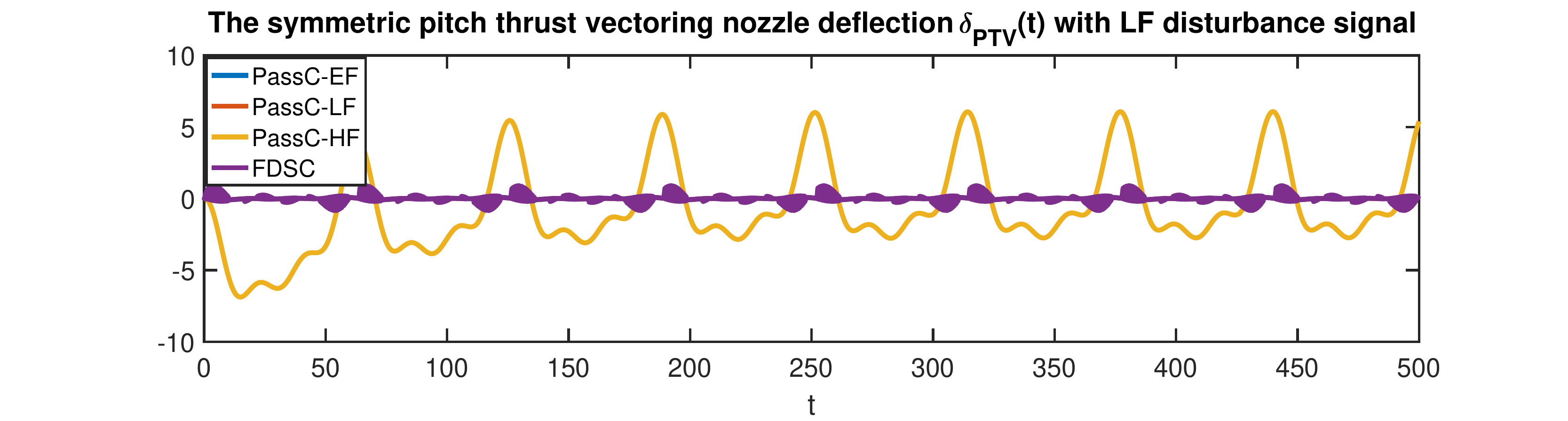}
            \caption{}
             \label{PTV-LF}
    \end{subfigure}
    \caption{The perturbation of the regulated output signal $z(t)=[q(t)~\delta_E(t)~\delta_{PTV}(t) ]^T$}
\end{figure}

In this example, we only use the LF and HF passive controllers (i.e., setting $K_1=K_{f_1}$ and $K_2=K_{f_3}$) to realize our proposed FDSC scheme. By solving the LMI optimization problem based upon the design conditions in Theorem \ref{Theorem 1}, the matrices $Q_1$ and $Q_2$ in the switching law can be computed as
\[\begin{array}{l}
	Q_{1} =10^6 \times\pmatset{1}{0.1pt}
	\pmatset{0}{0.1pt}
	\pmatset{2}{4pt}
	\pmatset{3}{4pt}
	\pmatset{4}{4pt}
	\pmatset{5}{4pt}
	\pmatset{6}{1pt} 
	\begin{pmat}[{.}]
		0.8559 & {\kern 2pt}  -0.2207\cr
		-0.2207   & {\kern 2pt} 0.0613\cr
	\end{pmat}, ~
	Q_{2} =\begin{pmat}[{.}]
		0.0211 &  {\kern 2pt} 0.0285\cr
		0.0285    &{\kern 2pt} 0.1449\cr
	\end{pmat}.
\end{array}\]
Let us consider two stationary cases with respect to the spectrum of disturbance.  To generate frequency-limited disturbance with LF and HF spectrum, a natural way is to set the disturbance as a sum of sinusoidal signals. For our first simulation, we set the disturbance as the LF disturbance
\begin{equation}\label{d-l}
	d(t)= \sin(0.1t)+\sin(0.2t)+\sin(0.3t)
\end{equation}
with spectrum $\Omega_1$. Fig. \ref{q-LF}-Fig. \ref{PTV-LF} show the perturbation pitch rate $q(t)$, symmetric horizontal tail deflection $\delta_{E}(t)$ and symmetric pitch thrust vectoring nozzle deflection $\delta_{PTV}(t)$ of the closed-loop system with PassC-EF, PassC-LF, PassC-HF and FDSC, as the simulation results. From Fig. \ref{q-LF}-Fig. \ref{PTV-LF}, we can notice that the fluctuation range of the red line is close to zero, which means the anti-disturbance performance of PassC-LF is the best. On the contrary, the disturbance attenuation performance of PassC-HF is the worst, because the fluctuation range of the yellow line is up to a thousand times greater than for the others. Although the fluctuation range of the purple line is not the smallest, the gap between the purple and red lines is far less than the gap between the purple and yellow lines. In case of LF signal input, our proposed FDSC scheme can provide pretty good anti-disturbance performance that is similar to PassC-LF. 

\begin{figure}[t!]
    \centering
    \begin{subfigure}[b]{0.9\textwidth}
           \centering
           \includegraphics[width=\textwidth]{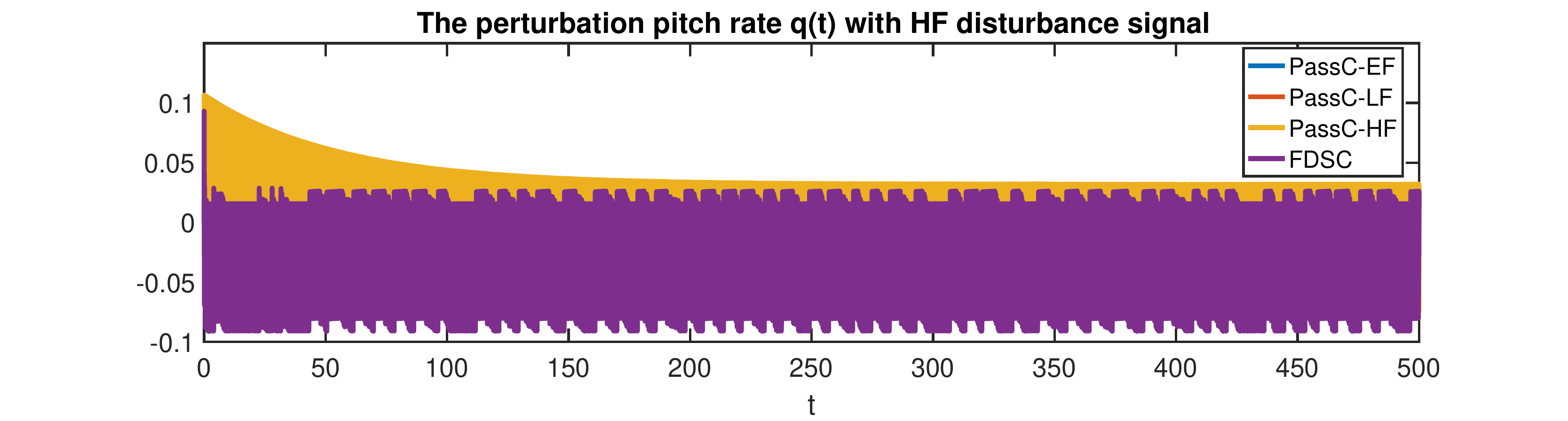}
            \caption{}
             \label{q-HF}
    \end{subfigure}
    \begin{subfigure}[b]{0.9\textwidth}
            \centering
            \includegraphics[width=\textwidth]{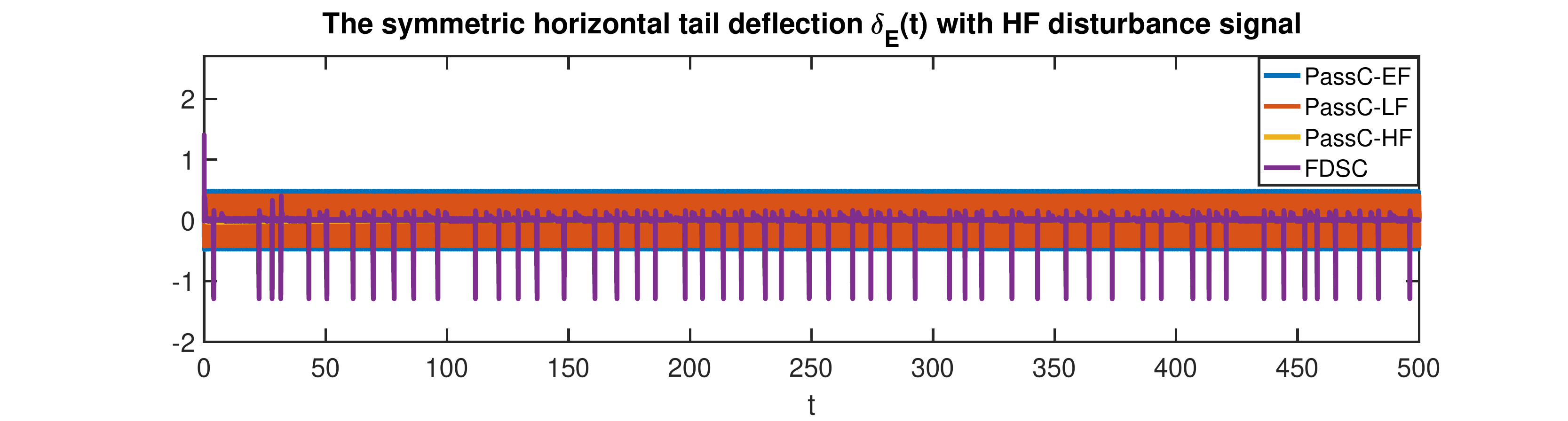}
            \caption{}
             \label{E-HF}
    \end{subfigure}
    \begin{subfigure}[b]{0.9\textwidth}
            \centering
            \includegraphics[width=\textwidth]{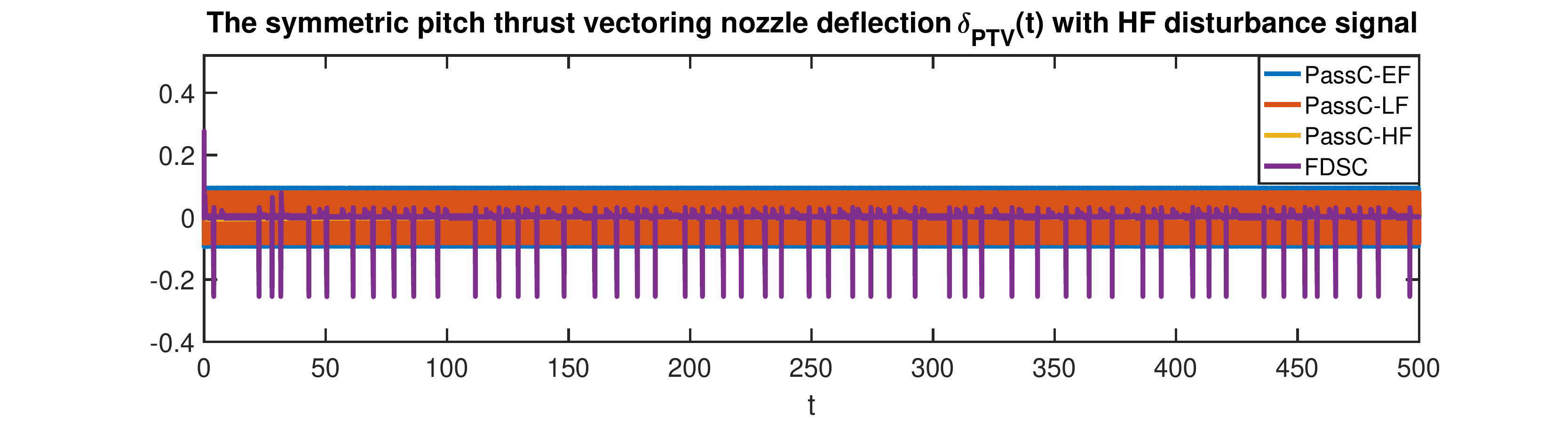}
            \caption{}
             \label{PTV-HF}
    \end{subfigure}
    \caption{The perturbation of the regulated output signal $z(t)=[q(t)~\delta_E(t)~\delta_{PTV}(t) ]^T$}
    \label{HF}
\end{figure}

\begin{figure}[!htbp]
	\centering
	\includegraphics[width=0.9\textwidth]{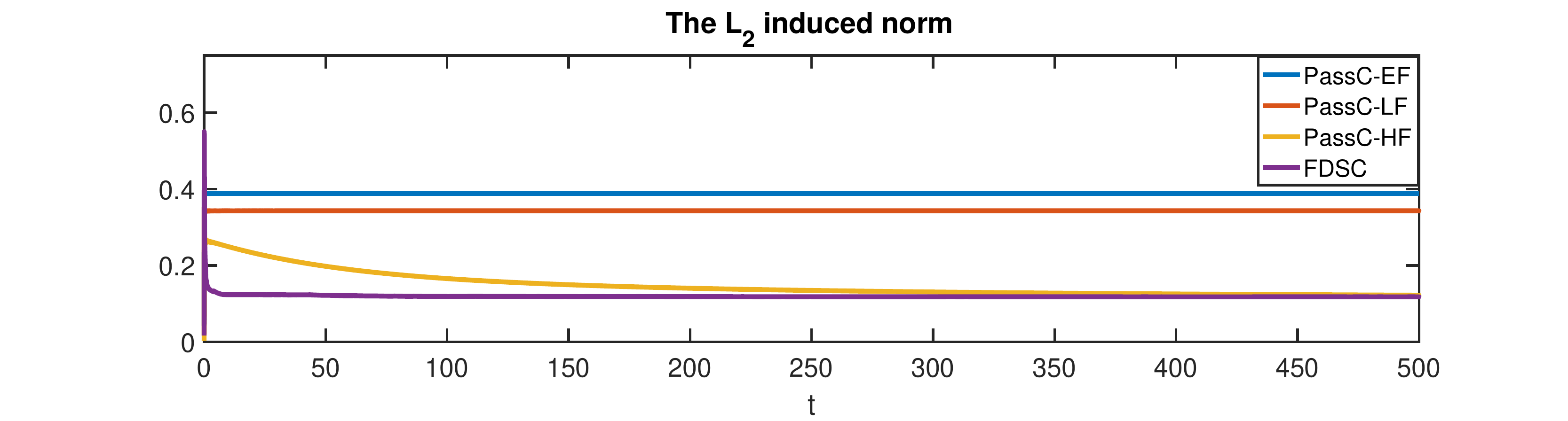}
	\caption{The $L_2$ induced norm between HF disturbance and regulated output signals}
	\label{gamma-HF}
\end{figure}

\noindent Similarly, we set the disturbance as the HF disturbance
\begin{equation}\label{d-h}
	d(t)=\sin(100t)+\sin(200t)+\sin(300t)
\end{equation}
with spectrum $\Omega_3$. Fig. \ref{HF} and Fig. \ref{gamma-HF} show the perturbation of the regulated output signal $z(t)$ and $L_2$ induced norm of the closed-loop system with PassC-EF, PassC-LF, PassC-HF and FDSC , as the simulation results. In Fig. \ref{q-HF}, the four lines are almost coincident and the fluctuation range of these lines is less than 0.15, while in Fig. \ref{gamma-HF}, the $L_2$ induced norm of PassC-EF and PassC-LF is nearly twice that of PassC-HF and FDSC which illustrates that the performance of FDSC is better than PassC-EF, PassC-LF and close to PassC-HF in this scenario. It is easy to see that the gap between the purple and yellow lines is far less than the gap between the purple and red lines from Fig. \ref{E-HF}-Fig. \ref{PTV-HF}. In case of an HF signal input, FDSC can still provide excellent disturbance attenuation performance which is similar to PassC-HF.

To test the disturbance attenuation capability of the passive controllers and our FDSC schemes, we first consider the scenario that an assumed HF disturbance is corrupted by some LF components, i.e.,
\[\begin{array}{l}\label{d-add}
	d(t)=d_h(t)+\rho_{p} d_l(t) \\
	{\kern 19pt}=\sin(100t)+\sin(200t)+\sin(300t) +\rho_{p} (\sin(0.1t)+\sin(0.2t)+\sin(0.3t)),
\end{array}\]
where $\rho_{p}$ represents the relative power intensity of the mixed out-of-band LF components. We set the simulation time to $t\in [0,500]$, and choose different $\rho_{p}$ over [0, 0.5]. For a clear illustration, we repeat the simulation with respect to different chosen $\rho_{p}$.

\begin{figure}[!htbp]
	\centering
	\includegraphics[width=0.9\textwidth]{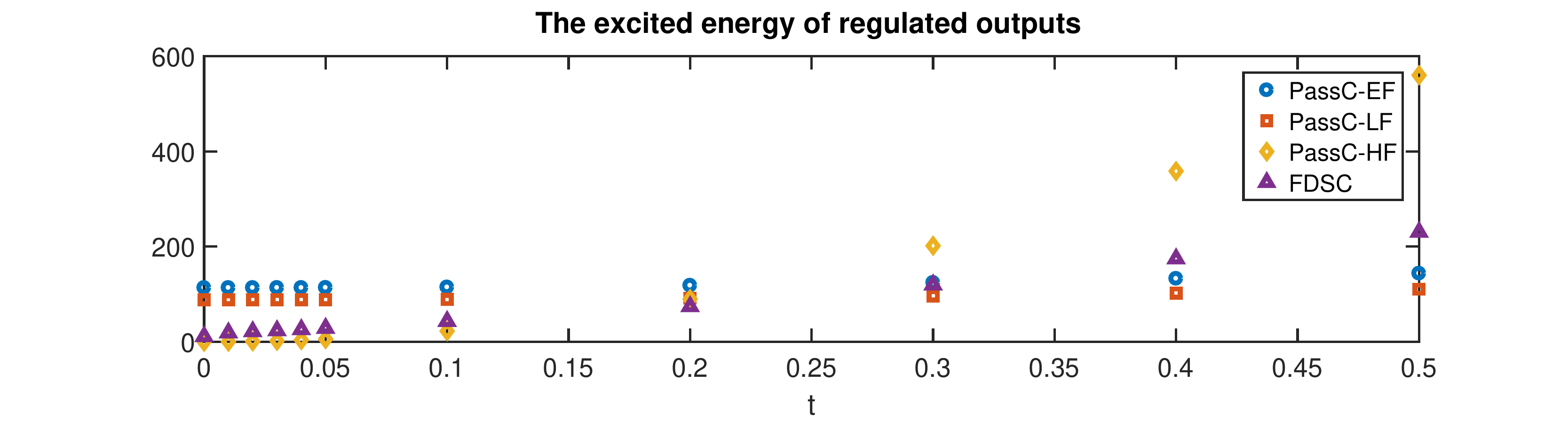}
	\caption{The excited energy of regulated outputs over $t\in [0,500]$ with different $\rho_p$}
	\label{power-para}
\end{figure}

It should be noted that since the energy of PassC-HF has reached $10^6$, there is a big difference with the other controllers, hence we use $1/5000$ energy value of PassC-HF to compare with others. Fig. \ref{power-para} shows the excited energy of the regulated outputs with respect to different controllers. It is seen that the PassC-HF controller is very sensitive to the mixed out-of-band (LF) components, using it actually becomes unacceptable even if the mixed LF out-of-band components are very small. Nevertheless, the proposed FDSC gives rise to excellent balanced performance in case that the out-of-band components are relative small ($\rho_{p}\leq 0.2$). Certainly, as expected,  in case that the out-of-band components are of comparatively large magnitude ($\rho_{p}> 0.2$), using the standard EF $H_\infty$ controller or PassC-LF maybe a better option.  

Next, we consider the scenario that the assumed HF disturbance is varied in a frequency-spectrum fluctuation manner, i.e., we insert an out-of-band LF signal in the running time of the in-band HF signal, i.e., the disturbance is expressed in sum of sinusoidal signals
\begin{equation}\label{d-piece}
	d(t)= \left\{ \begin{array}{l}
		\sin(100t)+\sin(200t)+\sin(300t),  {\kern 40pt}     \forall t \in [0,T),\\
		\rho_p^*(\sin(0.1t)+\sin(0.2t)+\sin(0.3t)),{\kern 28pt}  \forall t \in  [T, T+\rho_tT),\\
		\sin(100t)+\sin(200t)+\sin(300t),  {\kern 40pt}     \forall t \in [T+\rho_tT,2T+\rho_tT),\\
	\end{array} \right.
\end{equation}
where $\rho_t\in[0,0.5]$ is the radio of the embedded. We set $T=500$ and $\rho_p^*=0.1$ in the following group of simulations. 

\begin{figure}[!htbp]
	\centering
	\includegraphics[width=0.9\textwidth]{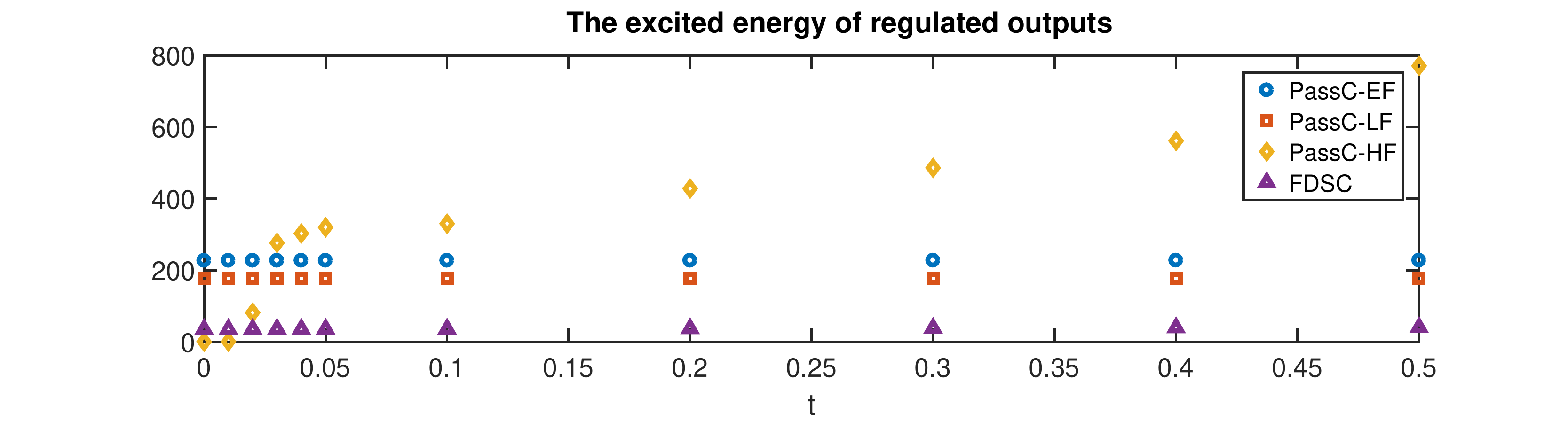}
	\caption{The excited energy of regulated outputs over $t\in [0,1000+500\rho_t]$ with different $\rho_t$}
	\label{D-LF}
\end{figure}
\noindent Due to a big energy value gap between PassC-HF and other controllers, $1/100$ of the energy value of PassC-HF is used to compare with other controllers. In Fig. \ref{D-LF}, we see a severe deterioration of the PassC-HF disturbance attenuation performance-even the embedded out-of band (LF) components persist in a very short time window, but the proposed FDSC technique displays excellent in rejecting disturbances of the system, unsurprisingly. With the increase of the ratio of out-of-band components, the banded dominance characterization of the frequency spectrum becomes more and more vague, it is hard to consider $d(t)$ is a HF-dominated signal. Under such a circumstance, using the standard EF $H_\infty$ controller or PassC-LF maybe a better option.

\section{Conclusion}
We have revealed the frequency-selectiveness property of a class of frequency-dependent functions with respect to the system state and its derivative for linear time-invariant systems. Based on this, we propose a frequency-dependent switching mechanism on the basis of a couple of or a group of frequency-selective passive controllers. An example illustrats the proposed FDSC control scheme and demonstrates its effectiveness on adjusting the in-band and out-of-band disturbance attenuation performance.

\begin{IEEEbiography}[{\includegraphics[width=1in,height=1.25in,clip,keepaspectratio]{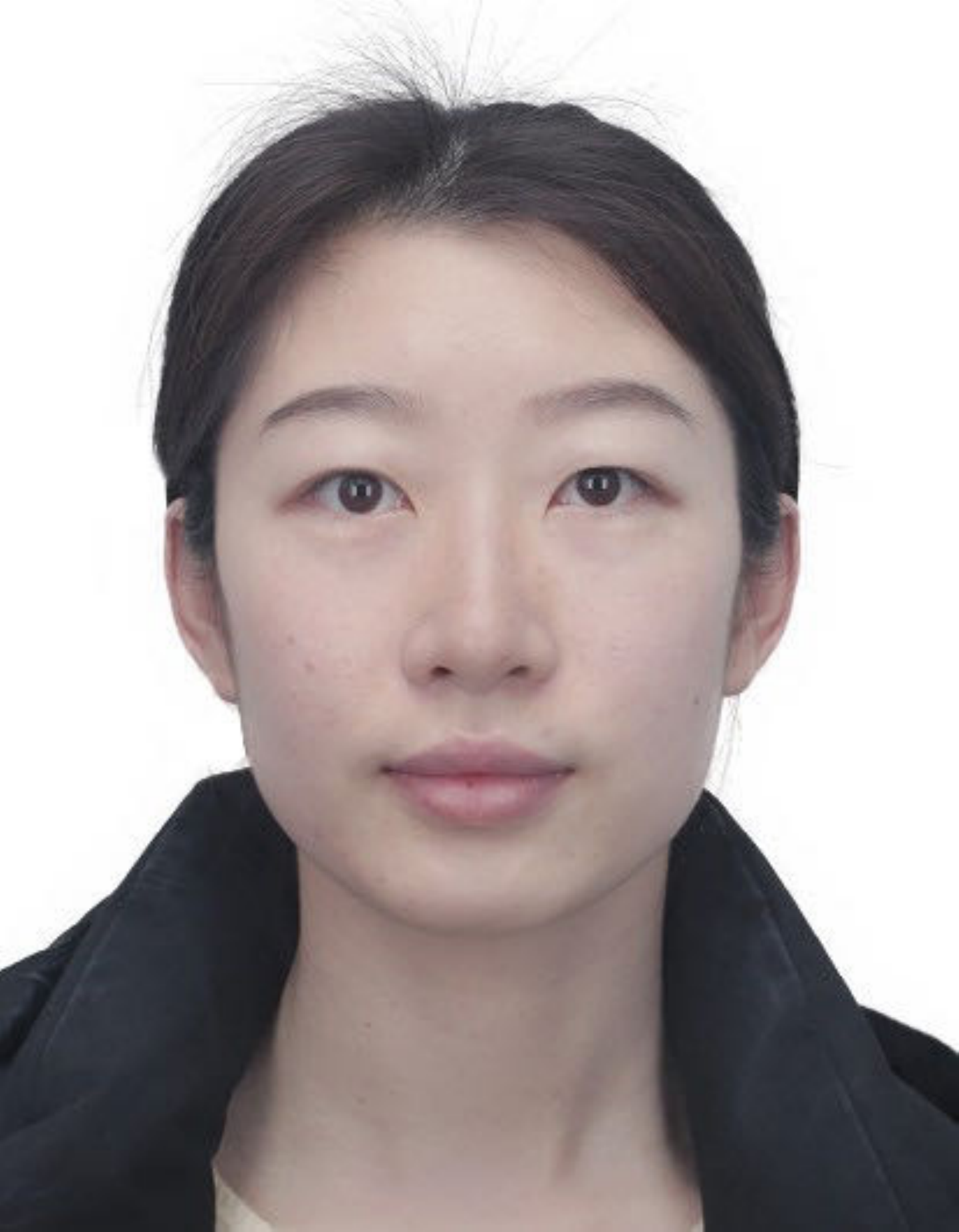}}]{Jingjing Zhang} (M'94) received the B.S. degree in mathematics and applied mathematics from Hebei Normal University, Hebei, China, in 2017 and the M.S. degree in basic mathematics from Liaoning University, Liaoning, China, in 2020. 	She is currently pursuing the Ph.D. degree in control theory and control engineering at Shanghai University, Shanghai, China. 
	
From 2021 to now, she was a joint Ph.D student in department computational methods in systems and control theory (CSC) at Max Planck Institute for Dynamics of Complex Technical Systems, Magdeburg, Germany. Her research interest includes robust control, finite frequency switching control and adaptive control.
\end{IEEEbiography}

\begin{IEEEbiography}[{\includegraphics[width=1in,height=1.25in,clip,keepaspectratio]{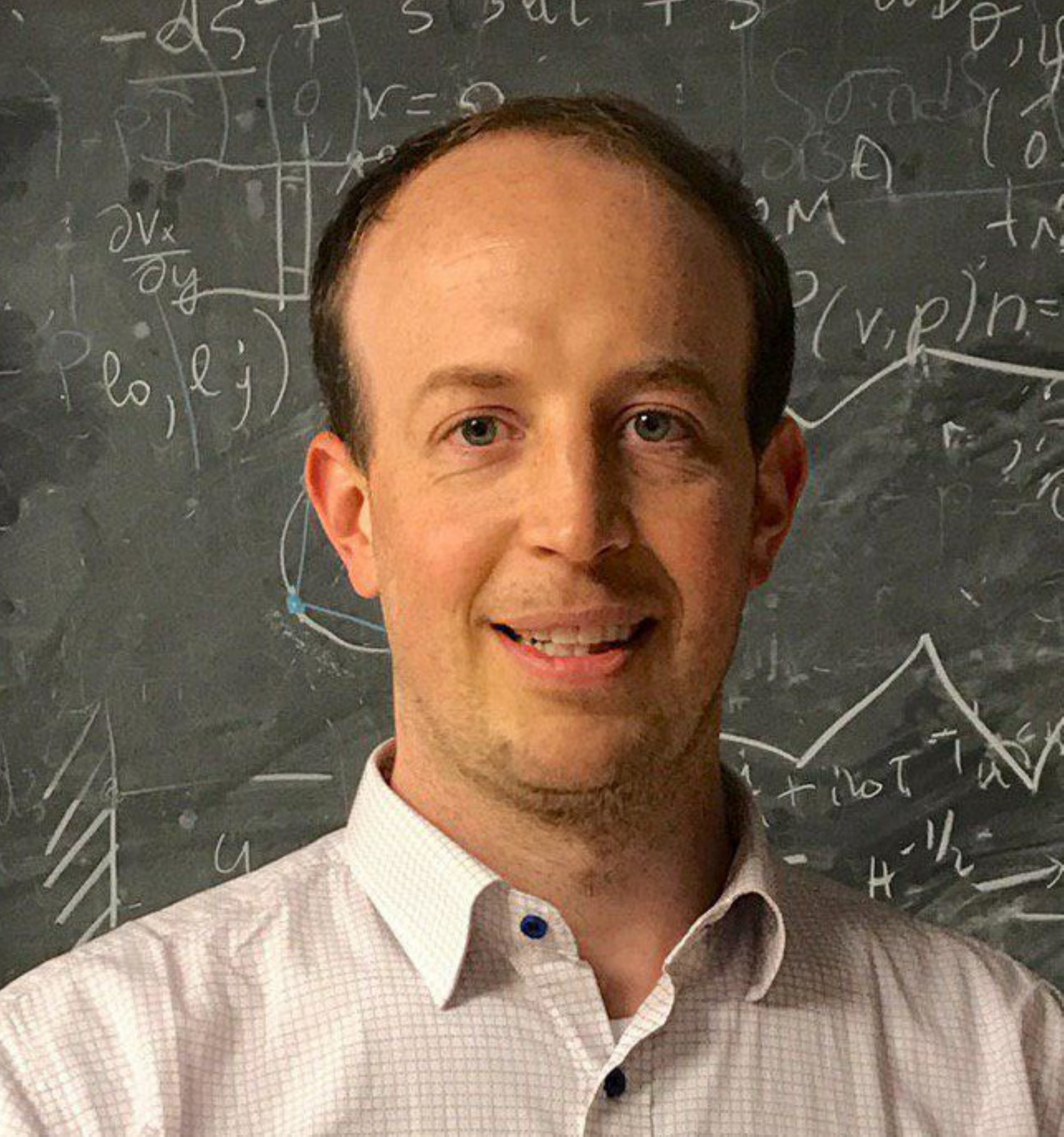}}]{Jan Heiland} received the Ph.D degree in mathematics from the Technical University of Berlin in 2013. Jan Heiland is a researcher and team leader at the Max Planck Institute for dynamics of complex technical systems in Magdeburg, Germany. He has gathered two years of practical experience in the private sector dealing with high-speed trains like the Talgo. Also, he has teaching duties at the University of Magdeburg. During February and March 2020 he collaborated as Visiting Professor at the ERC Advanced Grant project DyCon with Prof. Enrique Zuazua (FAU, University of Deusto and Universidad Autónoma de Madrid).
\end{IEEEbiography}

\begin{IEEEbiography}[{\includegraphics[width=1in,height=1.25in,clip,keepaspectratio]{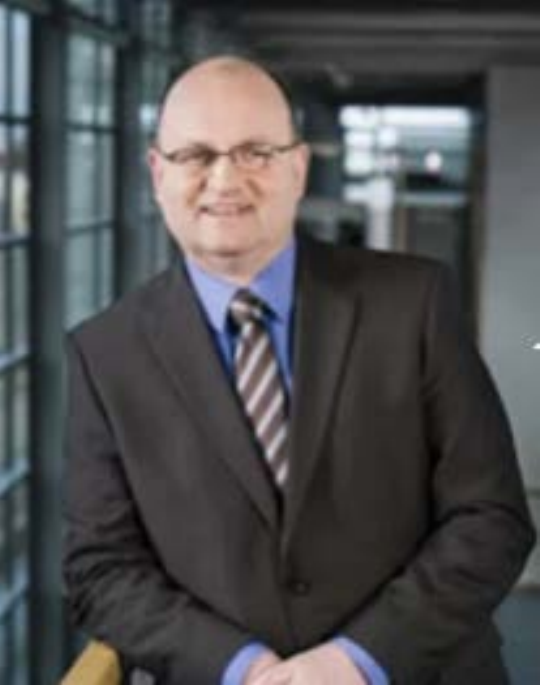}}]{Peter Benner} received the Diploma degree in mathematics from the RWTH Aachen University, Aachen, Germany, in 1993, the Ph.D. degree in mathematics from the University of Kansas, Lawrence, KS, USA, and the TU Chemnitz-Zwickau, Germany, in February 1997, and the Habilitation (Venia Legendi) degree in mathematics from the University of Bremen, Germany, in 2001. After spending a term as a Visiting Associate Professor with the TU Hamburg-Hamburg, Germany, he was a Lecturer in Mathematics with the TU Berlin, Germany, from 2001 to 2003. Since 2003, he has been a Professor of mathematics in industry and technology with the TU Chemnitz. In 2010, he was appointed as one of the four Directors of the Max Planck Institute for Dynamics of Complex Technical Systems, Magdeburg, Germany. Since 2011, he has been an Honorary Professor with the Otto-von-Guericke University of Magdeburg, Germany. His research interests include scientific computing, numerical mathematics, systems theory, and optimal control. He is a SIAM Fellow (Class of 2017).
\end{IEEEbiography}

\begin{IEEEbiography}[{\includegraphics[width=1in,height=1.25in,clip,keepaspectratio]{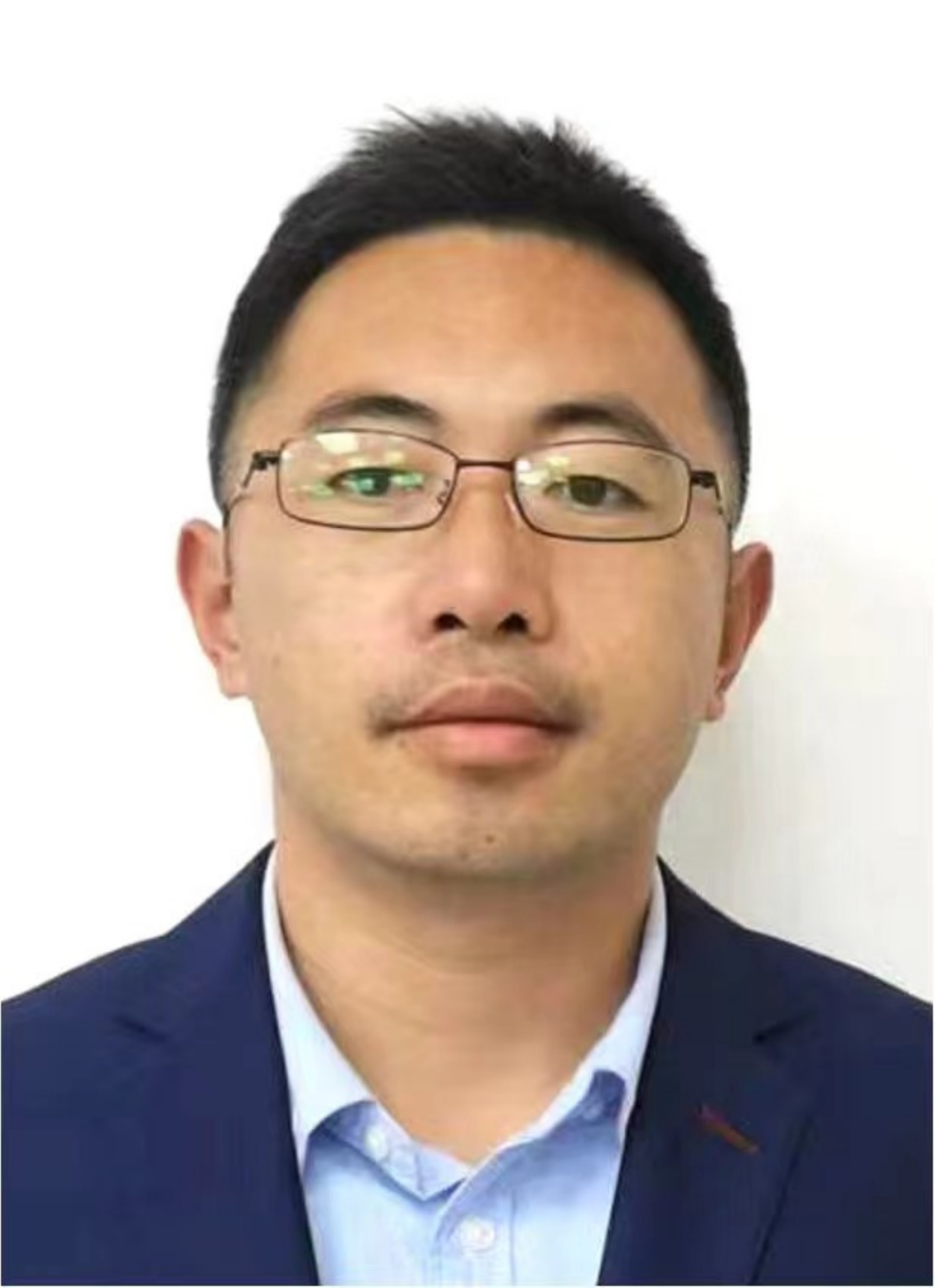}}]{Xin Du}received his B.S degree from University of Science and Technology Beijing, Beijing, China, in 2004. He received Ph.D. degree in Control Theory and Control Engineering from Northeastern University, Shenyang, China, in 2010. Since February 2010, He joined Shanghai University from February 2010, where he is currently an Associate Professor. He has been a Post-doc research fellow at Max-Planck Institute for Dynamics of Complex Technical Systems, Magdeburg, Germany from April to August, 2012 and from December 2013 to August 2015. During March 2016 to February 2018, he served as a Program Manager of Bureau of International Cooperation, National Natural Science Foundation of China. His research interests include modeling and control of complex systems, model order reduction, dynamical system and machine learning.  
\end{IEEEbiography}

\end{document}